\documentclass[aps,preprint,showpacs,preprintnumbers,amsmath,amssymb]{revtex4}

\usepackage{graphicx}

\newcounter{saveeqn}
\newcommand{\alpheqn}{\setcounter{saveeqn}{\value{equation}}%
  \stepcounter{saveeqn}\setcounter{equation}{0}%
  \renewcommand{\theequation}{%
  \mbox{\arabic{saveeqn}\alph{equation}}}}%
\newcommand{\reseteqn}{\setcounter{equation}{\value{saveeqn}}%
\renewcommand{\theequation}{\arabic{equation}}}

\begin{document}

\title{General variational many-body theory with complete self-consistency for trapped bosonic systems}
\author{Alexej I.\ Streltsov\footnote{E-mail: alexej@tc.pci.uni-heidelberg.de}}
\author{Ofir E.\ Alon\footnote{E-mail: ofir@tc.pci.uni-heidelberg.de}} 
\author{Lorenz S.\ Cederbaum\footnote{E-mail: Lorenz.Cederbaum@urz.uni-heidelberg.de}} 
\affiliation{Theoretische Chemie, Universit\"at Heidelberg, D-69120 Heidelberg, Germany}
\date{\today}
\begin{abstract}
In this work we develop a complete variational many-body theory for a system of $N$ trapped bosons
interacting via a general two-body potential. The many-body solution of this system
is expanded over orthogonal many-body basis functions (configurations).  
In this theory both the many-body basis functions {\em and} the respective expansion coefficients 
are treated as variational parameters.
The optimal variational parameters are obtained {\em self-consistently} by solving 
a coupled system of non-eigenvalue -- generally integro-differential -- equations 
to get the one-particle functions and
by diagonalizing the secular matrix problem to find the expansion coefficients.
We call this theory multi-configurational Hartree for bosons or MCHB(M),
where M specifies explicitly the number of one-particle functions used to construct 
the configurations. 
General rules for evaluating the matrix elements of 
one- and two-particle operators are derived and applied to construct the secular Hamiltonian matrix.
We discuss properties of the derived equations.
We show that in the limiting cases of one configuration the theory boils down to the 
well-known Gross-Pitaevskii and the recently developed multi-orbital mean-fields. 
The invariance of the complete solution with respect to unitary transformations
of the one-particle functions is utilized to find the solution with the minimal number of contributing
configurations.

In the second part of our work we implement and apply the developed theory. 
It is demonstrated that for any practical computation where 
the configurational space is restricted, the description of trapped bosonic systems strongly
depends on the choice of the many-body basis set used, i.e., self-consistency is of great relevance.
As illustrative examples we consider bosonic systems trapped in 
one- and two-dimensional symmetric and asymmetric double-well potentials.
We demonstrate that self-consistency has great impact on the predicted physical properties of 
the ground and excited states and show that the lack of self-consistency may lead to physically wrong 
predictions. The convergence of the general MCHB(M) scheme with a growing number M is validated 
in a specific case of two bosons trapped in a symmetric double-well.
\end{abstract}
\maketitle

\section{ Introduction }
The first experimental realizations of Bose-Einstein condensation 
in trapped ultra-cold atomic clouds \cite{1Exp1,1Exp2,1Exp3}
renewed a great interest in the experimental and theoretical descriptions of this phenomenon. 
Modern experimental setups utilize magnetic \cite{MagF}, electric \cite{EleF} and optical dipole fields
\cite{OptF} or their combinations to control the trapping and guiding of the ultra-cold atoms.
The number of condensed atoms in these experiments varies from several dozens \cite{Ndozen} to several 
millions \cite{1Exp1,1Exp2,1Exp3}.
Magnetically-tunable Feshbach resonances \cite{Feshb} make it possible to control the strength and sign of 
the inter-particle interactions.
All these experimental tools may be used to design a bosonic system \cite{Design1,Design2,Design3}
and to study its time-independent and time-dependent properties.

From the theoretical point of view we have to study the properties of the 
collection of interacting many-electron atoms immersed in a time-dependent crossing electric and magnetic fields.
This very complicated quantum mechanical many-body problem is replaced usually 
by a model Hamiltonian \cite{ModHam1,ModHam2,ModHam3,ModHam4}.
Typically, the diluteness of the atomic cloud allows to consider the atoms as point-like particles 
with pairwise interaction and to neglect three-body and higher order collisions.
The crossing electric and magnetic fields are replaced by an effective external trap potential.
Within these assumptions the original system is modelled as a collection of massive point-like particles 
interacting via repulsive or attractive pairwise inter-particle interaction 
immersed in an external trap potential of known geometry.

However, there are only a few examples of the model many-body Hamiltonians  
the exact solutions of which are known, see, e.g., Refs.~\cite{Exact1,Exact2,Exact3} and references therein.
For general inter-particle interactions and trap geometries the problem can be attacked only
within the framework of approximate and numerical methods. 
The most popular one is the variational approach,
where the form or a specific ansatz of the trial many-body function is postulated.
This ansatz depends on several parameters, and to solve the problem means to find 
the optimal set of the parameters which would minimize the expectation value of the Hamiltonian.
Clearly, the more parameters are involved 
the closer the obtained solution to the exact true many-body function is.

One of the widely and successfully used approximations is the Gross-Pitaevskii (GP) ansatz \cite{GP},
where it is assumed that each boson resides in a single spatial function, i.e.,
the many-body bosonic solution is presented as a product of 
identical one-particle functions (GP orbital):
\begin{equation}
\Psi(\vec{r}_1,\vec{r}_2,\ldots,\vec{r}_N)= \varphi(\vec{r}_1)\varphi(\vec{r}_2)\cdots\varphi(\vec{r}_N).
\label{ansatzGP}
\end{equation}
This ansatz has only one variational parameter - the shape of the GP orbital $\varphi$.
Orbitals with different shapes give different approximations to the true many-body solution.
The "optimal" orbital is obtained by minimizing the expectation value of the Hamiltonian with respect to $\varphi$,
which is equivalent to solving the very known GP equation.
The GP solution is {\it self-consistent}, i.e., the shape of the GP orbital
depends on the number of bosons and strength of the inter-particle interaction
in addition to the geometry of the external trap potential.
However, despite the great success of the GP ansatz, see Refs.\cite{ModHam1,ModHam2,ModHam3,ModHam4} 
and references therein,
there are ample physical situations this one-orbital mean-field theory cannot describe,
such as the depletion and fragmentation of trapped condensates and 
appearance of Mott-insulator phases of cold bosonic atoms in optical lattices.
The natural way to resolve these difficulties is to 
take a more general many-body ansatz and to go, therefore, beyond Gross-Pitaevskii theory.

Recently, a more general mean-field theory for bosons has been put forward \cite{BMF,BMFIMPL,ASCPL2005}. 
The multi-orbital, or best mean-field (BMF) theory as it is also called has been derived 
by considering the many-body ansatz where $n_1$ bosons reside in one orbital 
$\phi_1$, $n_2$ bosons in a second orbital $\phi_2$ and so on:
\begin{equation}
\Psi(\vec{r}_1,\ldots,\vec{r}_N)=
\hat{\cal S}\phi_1(\vec{r}_1)\cdots\phi_1(\vec{r}_{n_1})\phi_2(\vec{r}_{n_1+1})
\cdots\phi_2(\vec{r}_{n_1+n_2})\cdots \phi_M(\vec{r}_{n_1+n_2+\cdots+n_M}),
\label{ansatzBMF}
\end{equation}
where $\hat{\cal S} $ is the symmetrizing operator required for bosons.
In the multi-orbital mean-field theory the shapes of the one-particle 
functions $\phi_i$, the occupation numbers $n_i$ as well as the number $M$ of
one-particle functions are considered as variational parameters. 
The optimal parameters of this many-body function are 
obtained by solving a respective system of coupled non-linear equations. 
The resulting mean-field solution is also {\it self-consistent}, 
and depends on the number of bosons and strength of the inter-particle interaction
in addition to the geometry of the external trap potential.
The BMF theory was used, among several applications and predictions, to show \cite{Path} 
that with increasing inter-particle interaction,
the ground state of a trapped bosonic system on its pathway from condensation towards fermionization,
gradually passes through two-, three- and up to $M$-fold stages of the fragmentation.
The multi-orbital mean-field theory, when applied to the optical lattices, predicts \cite{Zoo}
the transition from the super-fluid to the Mott-insulator phase and reveals a variety of new 
Mott-Insulator phases.

The self-consistent GP and BMF theories considered above 
successfully describe the main features of the condensation, fragmentation and fermionization phenomena.
However, the mean-field ansatzes used are made of only one many-body function of known type 
(a permanent) and are, therefore, incapable -- by construction -- 
to describe the depletion and fluctuations of the many-body states.
To improve the many-body description further it is natural to go beyond 
mean-fields and take as an ansatz a linear combination of several {\it known} N-body basis functions:
\begin{equation}
|\Psi \rangle = \sum_{i} C_i | \Phi_i \rangle, 
\label{PSIexpans}
\end{equation}
where $\{| \Phi_i \rangle\}$ is a set of N-body basis functions and  $\{C_i\}$ are the expansion coefficients.
In quantum chemistry, Eq.(\ref{PSIexpans}) is called {\it configuration interaction} (CI) expansion \cite{CI},
because every many-body basis function ({\it configuration}) is attributed to
a simple physical situation. 
For bosonic systems, a configuration may represent a condensate, i.e., 
the state where all bosons reside in the same orbital, an excited state where 
$N-i$ bosons remain condensed and $i$ bosons are excited out of a condensate,
a two-fold fragmented state where a macroscopic number of bosons $n_1$ reside in one orbital 
and $n_2=N-n_1$ bosons in another, and so on.
The variational problem of finding the unknown expansion coefficients ${C_i}$ 
in this case is reduced to the diagonalization of the respective secular Hamiltonian matrix \cite{CI}.
This method is also known as exact diagonalization technique, because 
if all possible configurations are considered, i.e., the N-body basis is {\em complete}, 
then this expansion is {\em exact},
irrespective to the particular choice of the many-body basis functions $\{|\Phi_i\rangle\}$.
However, the secular Hamiltonian matrix in this case is an infinite matrix.

To make real computations tractable, the number of configurations in the expansion 
Eq.(\ref{PSIexpans}), of course, has to be truncated. The configurational space in this case
spans a restricted subspace of the Hilbert space.
The exact diagonalization studies within truncated configurational spaces 
have been used to provide a many-body description 
of repulsive and attractive condensates, see, e.g., Refs.\cite{Ueda,Schweds,FixOrb}.
In these investigations, 
many-body basis functions comprised of {\em a priori fixed orbitals} have been utilized
to study the properties of the bosonic system as a function of the inter-particle 
interaction strength. 
However, it is clear that exact diagonalizations 
performed within different truncations of the CI expansion will give different results.  
Moreover, truncated CI expansions of the same size but utilizing 
different sets of orbitals will also give different results in general.
Indeed, it was demonstrated \cite{Schweds,CCI} that the total energy obtained by the exact diagonalization 
in a restricted configurational space can sometimes be even worse than the self-consistent GP mean-field energy.
In these cases we encounter the situation that due to the implemented self-consistency, 
a single mean-field function provides a better description than a very large fixed-orbital many-body CI expansion. 

A question addressed in this work is how the choice
of the many-body basis functions impacts the results obtained within a restricted configurational space.
By comparing the many-body results obtained within different basis sets we can find 
the energetically most favorable one, i.e., the best many-body basis set.
The main goal of the present investigation is to formulate a many-body theory
which would provide the best set of many-body basis functions in 
a desired, i.e., truncated, subspace of the Hilbert space.
To achieve this goal, we apply a {\it general} variational principle where 
we treat {\it both} the set of many-body basis functions $\{| \Phi_i \rangle\}$ and 
the set of the expansion coefficients $\{C_i\}$
appearing in the expansion Eq.(\ref{PSIexpans}) as variational parameters which
are to be optimized. This results, as we shall see below, in a many-body theory for bosonic systems
with complete self-consistency, which we refer to as Multi-Configurational Hartree for Bosons or MCHB(M),
where M specifies explicitly the number of one-particle functions used to construct 
the configurations. 

Some hints that self-consistency is useful and important in attacking 
the many-boson problem {\it beyond} mean-field have already been addressed in the literature
in the context of symmetric double-well potentials and two-orbital configuration-interaction expansions. 
Spekkens and Sipe \cite{SpeSip} provide an approximative analytic as well as numerical solutions for the 
bosonic system trapped in symmetric double-well potentials
in the regime where the inter-particle interaction can be treated perturbatively.
More recently, Reinhard  \cite{Reinhard} and coworkers
have combined a partial nonlinear optimization of the many-body basis functions 
with a linear variational principle for the expansion coefficients to 
describe ground and excited states of bosons trapped in a symmetric double-well.  
Here, we would like to stress that the many-body variational theory which we develop in this paper
is {\em complete}, because we fully optimize all expansion coefficients as well as 
all the many-body basis functions appearing in the expansion Eq.(\ref{PSIexpans}), 
and {\em general} because it is valid for any trap geometry, 
for any physical shape of the inter-particle interaction and in any dimension.

The structure of the paper is as follows.
In Sec.\ref{SecII}, a trapped N-bosonic system interacting via a general 
pairwise potential is considered. 
The multi-configurational expansion Eq.(\ref{PSIexpans}) is used as an 
ansatz for the true many-body wave function of this system. 
The variational principle is used in Sec.\ref{SecIII} to formulate 
the general equations which allow one to find the best many-body basis functions 
and the corresponding expansion coefficients self-consistently in the desired active space.
Sec.\ref{SecIII.A} is devoted to the problem of finding the expansion coefficients.
In this section we provide the general rules for evaluating matrix elements 
of one- and two-body operators between two general basis functions (permanents)
and apply them to construct the secular Hamiltonian matrix. 
The problem of finding the optimal basis functions is considered  
in Sec.\ref{SecIII.B}, where  
the working equations for the general case are derived in closed form
using the elements of the reduced one- and two-body density matrices.
Properties of the resulting equations are discussed in Sec.\ref{SecIII.C}.
In particular, we demonstrate that in the limiting cases of a single permanent, the derived equations boil down to 
the one-orbital Gross-Pitaevskii and multi-orbital mean-fields equations.
Sec.\ref{SecIV.I} opens the second part of our work and provides explicitly the working equations
of MCHB(2), i.e., the case where
the basis functions are obtained as all possible permutations of N bosons over two orbitals.
As an inter-particle potential we implement the popular contact interaction.
The formalism is used to investigate the impact of self-consistency on many-body predictions.
The ground state of N=1000 bosons trapped in a symmetric 
one-dimensional double-well trap is investigated   
in Sec.\ref{SecIV.II} and for an asymmetric one in Sec.\ref{SecIV.III}.
In Sec.\ref{SecIV.IV} we show for these examples 
that the number of many-body basis functions
contributing to the expansion in Eq.(\ref{PSIexpans}) can be significantly reduced
by appropriately choosing the orbitals used to construct the many-body basis functions.
Excited states of the bosonic system trapped in a symmetric double-well trap
are investigated in Sec.\ref{SecIV.V}.
The two-dimensional bosonic system trapped in symmetric double-well is studied in Sec.\ref{SecIV.VI}.
The convergence of the MCHB(M) theory is addressed 
in Sec.\ref{SecV} where we apply different levels M$=2,\cdots,10$ of MCHB(M) theory to study
ground state properties of two bosons trapped in a symmetric double-well trap.
Finally, Sec.\ref{SecVI} summarizes our results and conclusions.

\section{Theory}

\subsection{Preliminaries}\label{SecII}
Consider a system of $N$ identical spinless bosons of mass $m$ 
immersed in an external time-independent  trap potential $V(\vec{r})$ and
interacting via a general pairwise interaction potential $W(\vec{r}_i-\vec{r}_j)$ 
where $\vec{r}_i$ is the position of the i-th particle.
By using bosonic annihilation and creation operators 
$b_i$ and $b_j^\dagger$,  $b_ib_j^\dagger-b_j^\dagger b_i=\delta_{ij}$
which are associated with a given set of orbitals $\{\phi_i\}$,
the Hamiltonian of the system takes on the standard form in second quantization language:
\begin{eqnarray}
\hat{H}&=&\hat h + \hat W \nonumber\\ 
\hat{h}=\sum_{i,j=1}h_{ij}b_i^\dagger b_j &,&
\hat{W}=\frac{1}{2}\sum_{i,j,k,l=1}W_{ijkl} b_i^\dagger b_j^\dagger b_l b_k,
\label{hamiltonianSQ}
\end{eqnarray}
where the one- and two-body matrix elements read
\begin{equation}
h_{ij}=\int \phi_i^*(\vec{r})(-\frac{\hbar^2}{2m}\nabla^2_{\vec{r}}
+ V(\vec{r})) \phi_j(\vec{r}) d\vec{r}
\label{Ih}
\end{equation}
and 
\begin{equation}
W_{ijkl}=\int\int\phi_i^*(\vec{r})\phi^*_j(\vec{r}')W(\vec{r}-\vec{r}')
\phi_k(\vec{r})\phi_l(\vec{r}')d\vec{r}d\vec{r}'.
\label{Iw}
\end{equation}

Our general intention is to find time-independent solutions of this Hamiltonian in a form 
of expansion (\ref{PSIexpans}) over basis functions.
Every basis function being a many-body function must depend on the coordinates of all N bosons.
The simplest way to construct it is to take a product of different orthogonal orbitals,
so called Hartree product
$\Phi_i\equiv\Phi_i(\vec{r}_1,\vec{r}_2,\ldots,\vec{r}_N)=
\hat{\cal S}\phi_1(\vec{r}_1)\phi_2(\vec{r}_2)\cdots\phi_N(\vec{r}_N)$,
and to apply the symmetrizing operator $\hat{\cal S}$ to fulfil the Bose statistic.
When all the orbitals are identical, we obtain  
a GP-like many-body basis function (see Eq.(\ref{ansatzGP})) used to describe condensation.
If some fraction of bosons resides in one orbital and the rest in another one,  
we deal with a basis function describing two-fold fragmentation, and so on.
The general many-body basis function can be considered as one of the configurations 
resulting due to permutation of N bosons over M orbitals.
In second quantization language this general many-body function, also known as permanent, reads
\begin{equation}
\Phi_i(\vec{r}_1,\vec{r}_2,\ldots,\vec{r}_N)=
\frac{1}{\sqrt{n_1!n_2!n_3!\cdots n_M!}} (b_1^\dagger)^{n_1}(b_2^\dagger)^{n_2}
\cdots(b_M^\dagger)^{n_M}|vac\rangle=|n_1,n_2,n_3,\cdots,n_M\rangle
\label{PSISQ}
\end{equation}
where M is a number of the one-particle functions, $n_1+n_2+n_3+\cdots+n_M=N$, and $|vac\rangle$ is the vacuum.

We recall that if the many-body basis set ${| \Phi_i \rangle}$ in expansion (\ref{PSIexpans})
is {\it complete}, i.e., it spans the whole N-boson Hilbert space, then expansion (\ref{PSIexpans}) is exact 
irrespective to the particular choice of the basis functions used.
In the present formulation, this is achieved when the number of the one-particle functions $M\to\infty$.
However, to make real computations tractable the size of the expansion (\ref{PSIexpans}) and hence 
the number of the one-particle basis functions M has to be restricted.
Let us assume that the many-body basis set $\{|\Phi_i\rangle\}$ 
is formed by {\it all} possible configurations appearing as permutation of N bosons over M orbitals.
The total number of configurations, and therefore the size of the expansion in this case is 
$\left( \begin{array}{c} M+N-1 \\ N\\ \end{array} \right)$. 
For example, for a system of N=1000 bosons and M=3 orbitals the total number of configurations 
is 501,501 while for M=4, the number of configurations is already 167,668,501,
which would make practical computations impossible.
Another consequence of the truncation is that different sets of orbitals 
used to construct many-body basis sets of the same size would 
lead to different approximations to the exact many-body $\Psi$.
As truncated CI expansions become very demanding with increasing N and/or M, it is
of great advantage to exploit this property. Namely,
it is desirable to find not only 
expansion coefficients but also the "best" set of one-particle basis functions.
Combined together, these lead to the self-consistent optimal CI expansion.
To fulfil this goal, we apply in this work the variational principle, which
defines the energetically most favorable solution as the best one.

\subsection{The general variational approach}\label{SecIII}
Let us consider a system of N bosons and restrict the number of one-particle functions to M.
Then, the trial many-body function (expansion (\ref{PSIexpans})) takes on the following form:
\begin{equation}
|\Psi \rangle =\sum_{n_1,n_2,\cdots,n_M} C_{(n_1,n_2,n_3,\cdots,n_M)} |n_1,n_2,n_3,\cdots,n_M\rangle  
\equiv\sum_{\vec{n}}C_{\vec{n}}|\vec{n}\rangle,
\label{PSIexpansII}
\end{equation}
where $n_1,n_2,\cdots, n_M$ are the number of bosons residing in orbitals $\phi_1,\phi_2,\cdots\phi_M$.
The summation runs over all possible configurations 
$
\vec{n}=(n_1,n_2,n_3,\cdots,n_M),
$
preserving total number of particles $n_1+n_2+\cdots+n_M=N$.
The expectation value of the Hamiltonian evaluated with this trial function reads:
\begin{equation}
\langle \Psi|\hat{H}|\Psi \rangle= \sum_{\vec{n},\vec{n'}}
C^*_{\vec{n}} C_{\vec{n'}} \langle \vec{n}|\hat{H}|\vec{n'}\rangle.
\label{PHP}
\end{equation}
This expression depends on two types of variational parameters: 
on the expansion coefficients $\{ C_{\vec{n}} \equiv C_{(n_1,n_2,n_3,\cdots,n_M)}\}$ and 
on the particular choice of the one-particle functions $\phi_1,\phi_2,\cdots\phi_M\equiv\{\phi_i\}$. 
Our main goal is to find the values of these parameters 
for which $\langle\Psi|\hat{H}|\Psi\rangle $ is a minimum. The natural requirements on
normalization of the trial many-body function $\langle\Psi |\Psi \rangle=1 $ and orthonormalization 
of all the one-particle functions $\langle\phi_i|\phi_j\rangle=\delta_{ij}$  
allow us to formulate the minimization problem within Lagrange's method of undetermined multipliers.
Here we have to stress that these are the only constrains applied. 

The Lagrange energy functional takes on the form:
\begin{eqnarray}
{\cal L}[ \{C_{\vec{n}}\},\{\phi_i\} ] 
&=& 
\langle\Psi|\hat{H}|\Psi \rangle+{\cal E}(1- \sum_{\vec{n}}  C^*_{\vec{n}} C_{\vec{n}})
+
\sum_{i=1}^M \sum_{j=1}^M \mu_{ij}(\delta_{ij}-\langle\phi_i|\phi_j\rangle), 
\label{LagrangeF}
\end{eqnarray}
where ${\cal E}$ appears due to normalization constrain of the trial many-body function (\ref{PSIexpansII}), and 
$\mu_{ij}$ -- due to orthonormalization of all one-particle functions.
To find the minimum of this functional we put to zero all first derivatives of this
functional with respect to every $C^*_{(n_1,n_2,n_3,\cdots,n_M)}$
and $\phi^*_i$ appearing in the expansion Eq.(\ref{PSIexpansII}):
\alpheqn
\begin{eqnarray}
\frac{\partial{ {\cal L}[ \{C_{\vec{n}}\},\{\phi_i\}]}}{\partial C^*_{\vec{n}}}
=0= 
\sum\limits_{\vec{n}'}  C_{\vec{n}'} \langle \vec{n}|\hat{H}|\vec{n}'\rangle
 - {\cal E} C_{\vec{n}}, \label{LagrangeSCI} &\forall \vec{n}& \\
 \frac{\delta{{\cal L}[ \{C_{\vec{n}}\},\{\phi_i\} ]}}{\delta \phi^*_i}=0
= 
\frac{\delta \langle\Psi|\hat{H}|\Psi\rangle}{\delta \phi^*_i}-\mu_{i1}|\phi_1 \rangle- \mu_{i2}|\phi_2 \rangle
\cdots - \mu_{iM}|\phi_M \rangle, &i=1,\cdots,M. & \label{LagrangeSPSI}
\end{eqnarray}
\reseteqn
Here we used partial derivatives and functional derivatives to separate the variations with respect
to the expansion coefficients and the one-particle functions. 
Such a separation is permitted because $\mu_{ij}$ and $\phi_i$ do not depend explicitly on $C_{\vec{n}}$.
Eq.(\ref{LagrangeSCI}) defines the expansion coefficients 
when the set of the one-particle functions is given,
while Eq.(\ref{LagrangeSPSI}) finds the best, i.e., energetically most favorable one-particle 
functions when the set of the expansion coefficients $C_{\vec{n}}$ is known.

We can use the following strategy to solve the two-fold variational problem 
Eqs.(\ref{LagrangeSCI}) and (\ref{LagrangeSPSI}) {\it self-consistently}.
Starting from some guess for the orbitals $\{\phi_i\}$ we construct the initial many-body basis set. 
Then we use these initial fixed-orbital many-body basis functions to build up the secular Hamiltonian
matrix. As we shall see in the subsequent Sec.\ref{SecIII.A}, 
the first part of the variational principle Eq.(\ref{LagrangeSCI}), i.e., 
the problem of finding the unknown coefficients $C_{(n_1,n_2,n_3,\cdots,n_M)}$
can be reduced to the diagonalization of the secular Hamiltonian matrix.
The expansion coefficients obtained as the components of the respective eigenvector are utilized in 
the second part of the variational procedure Eq.(\ref{LagrangeSPSI}) which provides as its output 
a new approximation to the one-particle functions $\{\phi_i\}$, see for details Sec.\ref{SecIII.B}. 
This iterative scheme is repeated until convergence is achieved. 
We call the obtained optimal set of the one-particle
functions and respective expansion coefficients the self-consistent solution.

Since in this approach the many-body bosonic wave-function is presented as a sum of all possible 
configurations (symmetrized Hartree products) resulting as permutations of N bosons over M orbitals, 
we name this general variational method Multi-Configurational Hartree for Bosons or MCHB(M).
Here M specifies explicitly the number of one-particle functions involved. 

\subsubsection{Variations with respect to the expansion coefficients $\{C_{\vec{n}}\}$
          and general rules for evaluating matrix elements with permanents}\label{SecIII.A}
    
The quantities $\langle \vec{n}|\hat{H}|\vec{n}'\rangle$ 
appearing in Eq.(\ref{LagrangeSCI}) are the matrix elements 
of a matrix Hamiltonian ${\cal H}$. 
This system of equations can be written in matrix notations as
\begin{equation}
{\cal H} {\bf C}={\cal E} {\bf C},
\label{MCHB2MATR}
\end{equation}
where ${\bf C}$ is a column vector of the expansion coefficients $C_{\vec{n}}$.
As usual, the problem of finding the expansion coefficients $\{C_{\vec{n}}\}$ 
for a given set of one-particle functions $\{\phi_i\}$ boils down to an eigenvalue problem. 
This opens the possibility to attack not only the ground, but also excited states.

The main issue now is to evaluate the matrix elements 
$\langle \vec{n}|\hat{H}|\vec{n}'\rangle$ with permanents. 
As introduced in Eq.(\ref{hamiltonianSQ}), the Hamiltonian is made of an one-particle operator $\hat{h}$
and of a two-body inter-particle interaction operator $\hat{W}$.
It is useful to treat these one-body $\langle \vec{n}|\hat{h}|\vec{n}'\rangle$ and two-body 
$\langle \vec{n}|\hat{W}|\vec{n}'\rangle$ interaction terms individually.

In the following we report the general rules to evaluate the matrix elements of one-body
and two-body operators. For simplicity these operators are denoted $\hat{h}$ and $\hat{W}$
although they need not be the constituents of the Hamiltonian.
Let us distinguish between six generic types of permanents:
\begin{eqnarray}
|P_0\rangle&=&|n_1,n_2,\cdots,n_i,\cdots,n_j,\cdots,n_M\rangle 
\equiv|;n_i;n_j;\rangle \nonumber \\
|P_1\rangle&=&|n_1,n_2,\cdots,n_i+1,\cdots,n_j-1,\cdots,n_M\rangle 
\equiv|;n_i+1;n_j-1;\rangle \nonumber\\
|P_2\rangle&=&|n_1,n_2,\cdots,n_i+2,\cdots,n_j-2,\cdots,n_M\rangle 
\equiv|;n_i+2;n_j-2;\rangle \nonumber\\
|P_3\rangle&=&|n_1,n_2,\cdots,n_i+2,\cdots,n_j-1,\cdots,n_k-1,\cdots,n_M\rangle 
\equiv|;n_i+2;n_j-1;n_k-1;\rangle \nonumber\\
|P_4\rangle&=&|n_1,n_2,\cdots,n_i+1,\cdots,n_j+1,\cdots,n_k-2,\cdots,n_M\rangle 
\equiv|;n_i+1;n_j+1;n_k-2;\rangle \nonumber\\
|P_5\rangle&=&|n_1,n_2,\cdots,n_i+1,\cdots,n_j+1,\cdots,n_k-1,\cdots,n_l-1,\cdots,n_M\rangle \nonumber\\
 &\equiv&|;n_i+1;n_j+1;n_k-1;n_l-1;\rangle.
\label{P}
\end{eqnarray}
The permanent $|P_1\rangle$ can be obtained from $|P_0\rangle$ by excitation of a single boson from 
$\phi_j$ to $\phi_i$. The other four permanents, $|P_2\rangle$ to $|P_5\rangle$,
describe two-boson excitations of $|P_0\rangle$.
To shorten the notations we show only the occupation numbers of the involved orbitals.
For instance, the configuration $(;n_i-1;n_j+1;)$ differs from $\vec{n}\equiv(;n_i;n_j;)$ by
an excitation of a single boson from $\phi_i$ to $\phi_j$.
We stress that only if two permanents differ by the excitation of at most two bosons,  
the Hamiltonian matrix element evaluated with these permanents is non-zero.
All other permanents give vanishing matrix elements.
  
The matrix elements of an one-body operator $\hat{h}$ are non-zero if the
two permanents are either equal (diagonal contributions) or differ by an excitation of a single boson:
\begin{eqnarray}
\langle P_0 |\hat{h} |P_0\rangle&=&\langle ;n_i;n_j;|
\sum_{\alpha,\beta=1}^{M}h_{\alpha\beta}
b_{\alpha}^\dagger b_{\beta}
|;n_i;n_j; \rangle
 =\sum_{\alpha=1}^{M}h_{\alpha\alpha} n_{\alpha}, \nonumber \\
\langle P_1 |\hat{h} |P_0\rangle&=&
\langle ;n_i+1;n_j-1;| \hat{h} |;n_i;n_j; \rangle
=h_{ij} \sqrt{n_i+1} \sqrt{n_j}. 
\label{PH}
\end{eqnarray}
The diagonal matrix elements of a two-body operator $\hat{W}$ are:
\begin{eqnarray}
\langle P_0 |\hat{W} |P_0\rangle&=&
\langle ;n_i;n_j;|
\sum\limits_{\alpha,\beta,\gamma,\delta}\frac{1}{2}W_{\alpha\beta\gamma\delta} 
b_{\alpha}^\dagger b_{\beta}^\dagger b_{\gamma} b_{\delta}
|;n_i;n_j; \rangle \nonumber\\ 
&=&\frac{1}{2}\sum_{\alpha}^{M}n_{\alpha}(n_{\alpha}-1)
W_{\alpha\alpha\alpha\alpha}+ \frac{1}{2}\sum_{\beta \ne \alpha=1}^{M} n_{\alpha}n_{\beta}
(W_{\alpha\beta\alpha\beta}+W_{\alpha\beta\beta\alpha}). 
\label{PW0}
\end{eqnarray}
The matrix elements of a two-body operator $\hat{W}$ evaluated with 
permanents which differ by the excitation of one boson read:
\begin{eqnarray}
\langle P_1 |\hat{W} |P_0\rangle&=&
\langle;n_i+1;n_j-1;| \hat{W}
|;n_i;n_j; \rangle \nonumber\\ 
&=&\frac{1}{2}\sqrt{n_i+1}\sqrt{n_j}[\sum\limits_{\alpha=1, \alpha \ne i,j}^{M} n_{\alpha} 
(W_{i\alpha\alpha j}+W_{i\alpha j\alpha}) +
n_i W_{iiij}+ (n_j-1) W_{ijjj}].
\label{PW1}
\end{eqnarray}

The permanents which differ by the excitations of two bosons 
give the following matrix elements of the two-body operator $\hat{W}$:
\begin{eqnarray}
\langle P_2 |\hat{W} |P_0\rangle&=&
\langle;n_i+2;n_j-2;| \hat{W}
|;n_i;n_j; \rangle \nonumber\\ 
&=&\frac{1}{2}\sqrt{(n_j-1)n_j} \sqrt{(n_i+2)(n_i+1)} W_{iijj}, \nonumber\\
\langle P_3 |\hat{W} |P_0\rangle&=&
\langle;n_i+2;n_j-1;n_k-1;| \hat{W}
|;n_i;n_j; \rangle \nonumber\\ 
&=&\sqrt{(n_i+2)(n_i+1)} \sqrt{n_j n_k}W_{iijk}, \nonumber\\
\langle P_4 |\hat{W} |P_0\rangle&=&
\langle ;n_i+1;n_j+1;n_k-2;| \hat{W}
|;n_i;n_j; \rangle \nonumber\\ 
&=&\sqrt{n_k(n_k-1)} \sqrt{(n_i+1)(n_j+1)}W_{ijkk}, \nonumber\\
\langle P_5 |\hat{W} |P_0\rangle&=&
\langle;n_i+1;n_j+1;n_k-1;n_l-1;|\hat{W}
|;n_i;n_j; \rangle \nonumber\\ 
&=&\sqrt{(n_i+1)(n_j+1)} \sqrt{n_k n_l}(W_{ijkl}+W_{jikl}).
\label{PW2}
\end{eqnarray}

In summary, we have demonstrated that for any set of orthogonal one-particle functions $\{\phi_i\}$
used to construct the many-body basis functions, i.e., the permanents,
the variational problem of finding the unknown coefficients $C_{(n_1,n_2,n_3,\cdots,n_M)}$
is reduced to the diagonalization of the Hamiltonian secular matrix.
To construct the secular Hamiltonian matrix we have developed 
rules for evaluating the matrix elements of the $\hat{h}$ and $\hat{W}$ operators.
Due to the generality of the consideration, the developed rules 
are, of course, applicable for any one- and two-body operators.

\subsubsection{Variations with respect to the orbitals $\{\phi_i\}$ }\label{SecIII.B}

The functional differentiation of the energy functional Eq.(\ref{LagrangeF}) 
with respect to the one-particle functions $\phi_i^*$ 
results in a system of M coupled integro-differential equations
\begin{equation}
\frac{\delta \langle\Psi|\hat{H}|\Psi\rangle}{\delta \phi^*_i}=\mu_{i1}|\phi_1 \rangle+ \mu_{i2}|\phi_2 \rangle
\cdots + \mu_{iM}|\phi_M \rangle, \, i=1,\cdots,M.
\label{MCHB0}
\end{equation}
By solving these equations for given fixed values of the 
coefficients $C_{\vec{n}}$ one obtains the respective set of the 
one-particle functions $\{\phi_i\}$.

The main purpose of this section is to express these equations in an explicit form.
To fulfil this goal, we first re-write the expectation value of the Hamiltonian, Eq.(\ref{PHP}), in
a form where all the terms depending on orbitals are explicitly visualized and 
then apply functional differentiations.

The expectation value of the Hamiltonian can be rewritten in 
the following form:
\begin{equation}
\langle \Psi|\hat{H}|\Psi \rangle= \sum^M_{i,j}\rho_{ij} h_{ij}+
\frac{1}{2}\sum^M_{i,j,k,l}\rho_{ijkl} W_{ijkl}.
\label{EVHR}
\end{equation}
Here, $\rho_{ij}=\langle \Psi| b_i^\dagger b_j|\Psi \rangle$
are the elements of the reduced one-body density matrix:
$$
\rho(\vec{r}_1|\vec{r}'_1)=N\int \Psi^{*}(\vec{r}'_1, \vec{r}_2, \cdots 
\vec{r}_N) \Psi(\vec{r}_1,\vec{r}_2,\cdots, \vec{r}_N) d\vec{r}_2 d\vec{r}_3 \cdots d\vec{r}_N
$$
\begin{equation}
= \sum^M_{i,j} \rho_{ij} \phi^*_i(\vec{r}'_1) \phi_j(\vec{r}_1),
\label{DNS1}
\end{equation}
and $\rho_{ijkl}=\langle \Psi| b_i^\dagger  b_j^\dagger b_k b_l|\Psi \rangle$ 
are the elements of the two-body density matrix:
$$
\rho(\vec{r}_1,\vec{r}_2|\vec{r}'_1,\vec{r}'_2)=N(N-1)\int 
\Psi^{*}(\vec{r}'_1,\vec{r}'_2,\vec{r}_3,\cdots \vec{r}_N) \Psi(\vec{r}_1,\vec{r}_2,\vec{r}_3,\cdots,\vec{r}_N) 
d\vec{r}_3\cdots d\vec{r}_N
$$
\begin{equation}
= \sum^M_{i,j,k,l} \rho_{ijkl} \phi^*_i(\vec{r}'_1) \phi^*_j(\vec{r}'_2) \phi_k(\vec{r}_1) \phi_l(\vec{r}_2).
\label{DNS2}
\end{equation}
The matrix elements of the reduced one- and two-particle densities can be easily obtained:
$$
\rho_{ii}=\sum_{\vec{n}}C^*_{\vec{n}} C_{\vec{n}} n_i\equiv \langle \hat n_i  \rangle,
$$
\begin{equation}
\rho_{ij}=\sum_{\vec{n}}C^*_{\vec{n}} 
C_{(;n_i-1;n_j+1;)} \sqrt{n_i(n_j+1)},
\label{DNS1ij}
\end{equation}
$$
\rho_{iiii}=\sum_{\vec{n}}C^*_{\vec{n}} C_{\vec{n}}(n_i^2-n_i)\equiv 
\langle \hat n_i^2 \rangle -\langle \hat n_i\rangle,
$$
\begin{equation}
\rho_{ijij}=\sum_{\vec{n}}C^*_{\vec{n}} C_{\vec{n}} n_in_j\equiv \langle \hat n_i \hat n_j \rangle.
\label{DNS2ii}
\end{equation}
Here we present only the diagonal matrix elements of the two-particle density; 
the off-diagonal ones are collected in Appendix \ref{A1}.

Inspecting Eqs.(\ref{DNS1ij}) and (\ref{DNS2ii}) we see that the matrix elements of the 
one- and two-particle densities 
depend only on the expansion coefficients $C_{\vec{n}}$ and  
do not depend on the one-particle orbitals $\{\phi_i\}$ explicitly.
In other words, only the one- and two-body integrals $h_{ij}$, $W_{ijkl}$
appearing in Eq.(\ref{EVHR}) have a functional dependence on the one-particle functions $\phi^*_i$.
Hence, the functional differentiation has to be applied only to 
the integrals $h_{ij}$, $W_{ijkl}$ which 
we treat individually according to the general rules of functional differentiation:
\begin{equation}
\frac{\delta h_{ij}}{\delta \phi^*_i}=\frac{\delta \langle \phi^*_i|\hat{h} | \phi_j\rangle}{\delta \phi^*_i}
=\hat{h}| \phi_j\rangle,
\label{MCHBh}
\end{equation}

\begin{equation}
\frac{\delta W_{ijkl}}{\delta \phi^*_i}=\frac{\delta 
\langle \phi^*_i\phi^*_j|\hat{W}| \phi_k\phi_l\rangle}{\delta \phi^*_i}
=\hat{W}_{jl}|\phi_k\rangle , \, i\ne j,
\label{MCHBW}
\end{equation}
where we introduce the notation
\begin{equation}
\hat{W}_{jl}=\int \phi^*_j(\vec{r}') W(\vec{r}-\vec{r}')\phi_l(\vec{r}')d\vec{r}'
\label{Ow}
\end{equation}
for the {\it local} operators $ \hat{W}_{jl}$.

Using Eqs.(\ref{MCHBh},\ref{MCHBW},\ref{Ow}) the variation of the expectation value of the
Hamiltonian, Eq.(\ref{EVHR}), with respect to the $\phi^*_i$ takes on the following
very compact and appealing form
\begin{equation}
\frac{\delta \langle\Psi|\hat{H}|\Psi\rangle}{\delta \phi^*_i}=
 \sum^M_{j=1}[\rho_{ij} \hat{h}+
\sum^M_{k,l=1}\rho_{ikjl} \hat{W}_{kl}]|\phi_j\rangle.
\end{equation}
Finally, the functional differentiations of the Lagrange energy functional 
Eq.(\ref{LagrangeF}) with respect to the one-particle functions 
result in a system of $M$ coupled integro-differential equations: 
\begin{equation}
\sum^M_{j=1}[\rho_{ij} \hat{h}+
\sum^M_{k,l=1}\rho_{ikjl} \hat{W}_{kl}]|\phi_j\rangle
= \sum^M_{j=1}\mu_{ij}|\phi_j \rangle, \,
i=1,\cdots,M.
\label{MCHBM}
\end{equation}

The main result of the present section is as follows. For any given set of the expansion coefficients $C_{\vec{n}}$,
the best, i.e., energetically most favorable set of one-particle functions $\{\phi_i\}$ 
used to construct the many-body basis functions (permanents) is determined by solving Eq.(\ref{MCHBM}).

\subsection{Properties of the MCHB(M) equations}\label{SecIII.C}

In the formulation of MCHB(M) we assumed that the
expansion Eq.(\ref{PSIexpansII}) spans {\it all} possible configurations of $N$ bosons
over M one-particle functions. 
However, the derived equations for the optimal expansion coefficients (\ref{MCHB2MATR}) and 
orbitals (\ref{MCHBM}) are general and 
remain valid even in the case when 
the expansion Eq.(\ref{PSIexpansII}) is incomplete and comprises any limited subset  
of configurations.
Let us first consider the simplest limiting case, 
where the expansion Eq.(\ref{PSIexpansII}) contains only a single
permanent in which all $N$ bosons reside in one and the same orbital $\phi_1$:
\alpheqn
\begin{equation}
|\Psi \rangle = C_{(n_1)} |n_1\rangle.
\label{PSIGP}
\end{equation}
Of course, $n_1=N$ due to the conservation of the total number of particles
and $C_{(n_1)}\equiv1$ due to the normalization of $| \Psi \rangle$.
Then, the expectation value of the Hamiltonian Eq.(\ref{EVHR}) takes on the simple form
\begin{equation}
\langle\Psi|\hat{H}|\Psi\rangle=\rho_{11} h_{11}+ \frac{1}{2} \rho_{1111} W_{1111}
\end{equation}
and Eq.(\ref{MCHBW}) now reads
\begin{equation}
\left\{ \rho_{11}\hat{h}+ \rho_{1111} \hat{W}_{11} \right\}|\phi_1 \rangle= \mu_{11}|\phi_1 \rangle.
\end{equation}
The respective non-zero elements of the reduced one- and two-body density 
matrices Eqs.(\ref{DNS1ij},\ref{DNS2ii}) become trivial
$$
\rho_{11}=\langle \hat n_1 \rangle\equiv N,
$$
\begin{equation}
\rho_{1111}=\langle \hat n^2_1-\hat n_1 \rangle\equiv N(N-1).
\end{equation}
Obviously, for contact inter-particle interaction $W(\vec{r}-\vec{r}')=\lambda_0\delta(\vec{r}-\vec{r}')$
and putting $\phi_1\equiv \varphi$
we easily reproduce the famous GP mean-field energy functional  
\begin{equation}
E_{GP}=\langle\Psi|\hat{H}|\Psi\rangle=N [h_{11} + \frac{N-1}{2}\lambda_0\int\varphi(\vec{r})^4d\vec{r}]
\end{equation}
and the GP equation for the optimal orbital 
\begin{equation}
\left\{ -\frac{\hbar^2}{2m}\nabla^2_{\vec{r}}+ V(\vec{r}) +
\lambda_0 (N-1) |\varphi(\vec{r})|^2 \right\}|\varphi \rangle= \mu_{GP}|\varphi \rangle,
\end{equation}
\reseteqn
where $\mu_{GP}\equiv\mu_{11}/N$.

Next, let us demonstrate that 
in the more general one-configurational case, when the permanent is given by Eq.(\ref{ansatzBMF}),
the many-body MCHB(M) theory boils down to the multi-orbital BMF theory \cite{BMF,BMFIMPL,ASCPL2005}.
In this case the only permanent contributing to the expansion 
Eq.(\ref{PSIexpansII}) with $C_{(n_1,n_2,\cdots,n_M)}=1$
represents a configuration with $n_1$ bosons residing in $\phi_1$, $n_2$ in $\phi_2$, ...
$n_M$ in $\phi_M$.
Let us first rewrite the general expression for the total energy Eq.(\ref{EVHR})
in a form where the diagonal $E_{MF}$ and off-diagonal $E_{MB}$ contributions are separated:
\alpheqn
\begin{equation}
\langle \Psi|\hat{H}|\Psi \rangle= E_{MF}+E_{MB}
\end{equation}
where
\begin{equation}
E_{MF}=\sum^M_{i=1}[\rho_{ii} h_{ii}+\frac{1}{2}\rho_{iiii} W_{iiii}+
\frac{1}{2}\sum^M_{j=1,j\ne i}(\rho_{ijij}W_{ijij}+\rho_{ijji}W_{ijji})],
\end{equation}
and
\begin{equation}
E_{MB}=\sum^M_{i,j=1,i\ne j}(\rho_{ij} h_{ij} +
\frac{1}{2}\sum^M_{\{i,j,k,l\}} \rho_{ijkl}W_{ijkl}),
\end{equation}
where $\{i,j,k,l\}$ runs over all possibilities not included in the respective $E_{MF}$ part.
Since only one configuration forms the many-body expansion, 
only the diagonal matrix elements of the reduced one- and two-body density matrices given 
in Eqs.(\ref{DNS1ij},\ref{DNS2ii}) have non-zero values:
$$
\rho_{ii}\equiv\langle \hat n_i \rangle= n_i,
$$
$$
\rho_{iiii}\equiv\langle \hat n^2_i-\hat n_i \rangle= n_i^2-n_i,
$$
\begin{equation}
\rho_{ijij}=\rho_{ijji}\equiv\langle \hat n_i \hat n_j \rangle= n_i n_j.
\end{equation}
Consequently, $E_{MB}=0$ and only the $E_{MF}$ part of the total energy survives and boils down to
\begin{equation}
\langle \Psi|\hat{H}|\Psi \rangle=E_{MF}=
\sum^M_{i=1}n_i[h_{ii}+\frac{n_i-1}{2}W_{iiii}+\sum^M_{j=1,j\ne i}n_j(W_{ijij}+W_{ijji})].
\end{equation}
\reseteqn
Analogously, Eq.(\ref{MCHBM}) determining the optimal orbitals now reads:
\begin{equation}
[n_i \hat{h} + n_i(n_i-1)\hat{W}_{ii} + \sum^M_{j=1,j\ne i}n_i n_j \hat{W}_{jj}] |\phi_i\rangle
+\sum^M_{j=1,j\ne i}n_i n_j \hat{W}_{ji} |\phi_j\rangle 
 = \sum^M_{j=1}\mu_{ij}|\phi_j \rangle, \,
i=1,\cdots,M.
\label{MCHBMMF}
\end{equation}
These equations coincide with the multi-orbital mean-field equations, formulated and applied in 
Refs.\cite{BMF,BMFIMPL,ASCPL2005,Path,Zoo}. 

We conclude that in the limiting cases where only one permanent 
forms the many-body expansion Eq.(\ref{PSIexpansII}), 
the many-body MCHB(M) theory boils down to the respective mean-fields. 
On the other hand, if only one configuration in the many-body expansion is dominant, then 
the multi-orbital mean-field predictions are very close to the many-body ones.
Such a situation can be realized in real physical systems,
for example in deep multi-well traps.

When the expansion Eq.(\ref{PSIexpansII}) spans {\it all} possible configurations 
of N bosons over M one-particle functions, the MCHB(M) equations
possess some interesting and useful properties.
Suppose we solve this system of equations and find the optimal orbitals $\{\phi_i\}$ and 
corresponding expansion coefficients $\{C_{(n_1,n_2,n_3,\cdots,n_M)}\}$.
By applying an unitary transformation on the ${\phi_i}$
we can construct a new set of the one-particle functions 
and find the corresponding new set of the expansion coefficients $\{\tilde{C}_{(n_1,n_2,n_3,\cdots,n_M)}\}$.
This unitary transformation does not change (up to a phase factor)
the many-body wave-function Eq.(\ref{PSIexpansII}).
Consequently,
the total energy $\langle\Psi|\hat{H}|\Psi\rangle$ of the system is invariant with respect 
to any unitary transformation of the one-particle functions. 
Moreover, we can demonstrate that this is also valid at any iteration during 
the described procedure leading to the self-consistent solution of the MCHB(M) equations. 
The considered invariance of the equations
is explained by the fact that the expansion Eq.(\ref{PSIexpansII}) spans 
{\it all} possible configurations, i.e., it is {\it complete} within the provided 
subspace of one-particle functions.
Clearly, for an incomplete expansion this invariance is, in general, lost.
For instance, it was shown that the multi-orbital mean-field equations \cite{BMF,BMFIMPL,ASCPL2005}, 
which, as discussed above, can be considered as one-permanent MCHB(M) equations, indeed 
are in general not invariant with respect to unitary transformations of the orbitals $\{\phi_i\}$.

Having established the equivalence of the MCHB(M) solutions with respect to unitary transformations,
it is natural to find out which set of orbitals can provide additional physical insight.
Since the energy is invariant, we consider the reduced one- and two-particle density matrices.
Having at hand a reduced one-particle density matrix one can diagonalize it:
\begin{equation}
\rho(\vec{r}|\vec{r}')=\sum^M_{i} \rho_{i} \phi^{*{NO}}_i(\vec{r}') \phi^{NO}_i(\vec{r}).
\label{NOA}
\end{equation}
The eigenvectors $\phi^{NO}_i$ are referred to as {\it natural orbitals} and 
the eigenvalues $\rho_{i}$ of this matrix are the respective {\it occupation numbers}.
The natural occupation numbers $\rho_{i}$ can be considered 
as the average numbers of bosons residing in $\phi^{NO}_i$. 
The natural orbital analysis of the many-body solution is often used to characterize the system.
The system is condensed \cite{Penrose} when only a single natural orbital 
has a macroscopic occupation. If several natural orbitals have macroscopic occupation numbers 
then the system is called {\it fragmented} \cite{Nozieres}.
Of course, the natural orbital analysis can be applied to any state, also to excited states. Comparing
the natural orbitals and occupation numbers of the ground and excited states one can learn 
on the nature of the excited state.

Let us examine the properties of the natural orbitals in the MCHB(M) theory.
On the one hand, we know that each eigenvector of the one-body density 
matrix $\rho(\vec{r}|\vec{r}')$ can be expressed as a linear 
combination of the MCHB(M) orbitals.
On the other hand, we have seen that due to the completeness of the many-body expansion,
the MCHB(M) solution is invariant with respect to any unitary transformation applied.
We conclude that the natural orbitals themselves constitute a MCHB(M) solution as well. 
This finding gives us the freedom to characterize the MCHB(M) one-particle functions 
by natural orbitals. We shall do so in the following, unless explicitly mentioned.

Using the multi-configurational expansion Eq.(\ref{PSIexpans}) as an
ansatz for the true many-body wave function of a trapped bosonic system 
and applying the variational principle we derived general equations which allow one to find
the best many-body basis functions and the corresponding expansion coefficients self-consistently.
We have shown that these equations are also applicable in any desired active sub-space,
including the limiting case of one-permanent. In the latter case the derived equations boil down to 
the one-orbital Gross-Pitaevskii and multi-orbital mean-fields equations.
At this point it is very important to stress that all derived equations and 
conclusions are general, i.e., they are valid for any geometry of the external trap potential, 
for any physical shape of the inter-particle interaction and in any dimension. 

\section{Illustrative numerical examples and applications}\label{SecIV}

\subsection{Preliminaries and implementation of MCHB(2)}\label{SecIV.I}

In the present section we implement the developed MCHB(M) formalism for systems of cold bosonic atoms.
We consider the simplest case of M=2, i.e., MCHB(2) theory and apply 
it to the ground and excited states of the bosonic systems 
trapped in symmetric and asymmetric one- and two- dimensional double-well potentials.
While the ground state of the one-dimensional bosonic gas trapped 
in symmetric double-well potentials has been intensively studied 
in the literature see Refs.\cite{SpeSip,Reinhard,DW,DWQ} and references
therein, the many-body properties of bosonic systems 
at higher dimensions in symmetric and especially asymmetric double-well traps \cite{DWasymm} 
are open theoretical questions.

The configurational space of the MCHB(2) theory constitutes all symmetrized permutations 
of N bosons over two orbitals $\phi_1$ and $\phi_2$.
The trial many-body function Eq.(\ref{PSIexpans}) in this case is spanned by N+1 configurations: 
\begin{equation}
|\Psi \rangle  =\sum_{n_1=0}^N C_{(n_1,n_2)} |n_1,n_2\rangle
\label{MCHB2PSI}
\end{equation}
The condition $n_1+n_2=N$ implies that only one occupation number $n_1$ can be used to enumerate the
configurations.

The configurational space of MCHB(2) theory also implies that 
the non-zero Hamiltonian matrix elements can be only 
between the three generic permanents
\begin{equation}
|P_0\rangle=|n_1,n_2\rangle, \, |P_1\rangle=|n_1+1,n_2-1\rangle, \, |P_2\rangle=|n_1+2,n_2-2\rangle.
\label{P2}
\end{equation}
Using these permanents and the general rules 
for evaluating one-body and two-body matrix elements,
Eqs.(\ref{PH},\ref{PW0},\ref{PW1},\ref{PW2}),
the expectation value of the Hamiltonian (\ref{EVHR}) becomes 
\begin{eqnarray}
\langle\Psi|\hat{H}|\Psi\rangle&=&
\langle \hat n_1 \rangle h_{11} + \langle \hat n_2 \rangle h_{22} \nonumber\\
+\frac{ \langle \hat n_1^2 \rangle - \langle \hat n_1 \rangle}{2} W_{1111}
    &+&
\frac{ \langle \hat n_2^2 \rangle - \langle \hat n_2 \rangle}{2} W_{2222} + 
\langle \hat n_1\hat n_2 \rangle (W_{1212}+W_{1221}) \nonumber\\
    &+&
\rho_{12} h_{12} + \rho_{21} h_{21}+
\rho_{2111}W_{2111}+ \rho_{1112}W_{1112} \nonumber\\
    &+&
\rho_{2221}W_{2221} + \rho_{1222}W_{1222}
+\frac{\rho_{2211}}{2} W_{2211} +  \frac{\rho_{1122}}{2} W_{1122}.
\label{MCHB2H3}
\end{eqnarray}
The coupled system of integro-differential equations (\ref{MCHBM}) needed for determination of the
optimal one-particle functions reads:
\begin{eqnarray}
\left\{ \langle \hat n_1 \rangle\hat{h}+
(\langle \hat n_1^2 \rangle - \langle \hat n_1 \rangle) \hat{W}_{11} +
\langle \hat n_1\hat n_2 \rangle \hat{W}_{22} +
\rho_{2111} \hat{W}_{21} + \rho_{1112}\hat{W}_{12} \right\}|\phi_1\rangle &+& \nonumber\\
\left\{ \langle \hat n_1\hat n_2 \rangle \hat{W}_{21} + \rho_{1112}\hat{W}_{11}+
\rho_{12}\hat{h}+ \rho_{1222}\hat{W}_{22}+\rho_{1122}\hat{W}_{12}
\right\}|\phi_2\rangle &=& \mu_{11}|\phi_1 \rangle+ \mu_{12}|\phi_2 \rangle \nonumber\\
\left\{ \langle \hat n_2 \rangle\hat{h}+
(\langle \hat n_2^2 \rangle - \langle \hat n_2 \rangle) \hat{W}_{22} +
 \langle \hat n_1\hat n_2 \rangle \hat{W}_{11} +
\rho_{2221}\hat{W}_{21} + \rho_{1222} \hat{W}_{12} \right\}|\phi_2\rangle &+& \nonumber\\
\left\{ \langle \hat n_1\hat n_2 \rangle \hat{W}_{12}+ \rho_{2221}\hat{W}_{22}+
\rho_{21}\hat{h}+\rho_{2111} \hat{W}_{11}+\rho_{2211}\hat{W}_{21}
\right\}|\phi_1\rangle &=& \mu_{21}|\phi_1 \rangle+ \mu_{22}|\phi_2 \rangle.  \nonumber\\
\label{MCHB2DLDPHI1}
\end{eqnarray}

As an illustrative example we consider N identical spinless bosons,
interacting via contact potential $W(\vec{r}-\vec{r}')=\lambda_0\delta(\vec{r}-\vec{r}')$,
where $\lambda_0$ is proportional to the $s$-wave scattering length.
In this case the two-body integrals, see Eq.(\ref{Iw}), appearing in Eq.(\ref{MCHB2H3}) simplify to
$$
W_{ijkl}=
\lambda_0\int\phi^*_i(\vec{r})\phi^*_j(\vec{r})\phi_k(\vec{r})\phi_l(\vec{r})d\vec{r},
$$
and instead of the local integral operators appearing in Eq.(\ref{MCHB2DLDPHI1}) we obtain the 
functions
$$
\hat{W}_{ij}=
\lambda_0\phi^*_i(\vec{r})\phi_j(\vec{r}).
$$
This considerably simplifies the implementation of the MCHB theory as 
the system of integro-differential equations (\ref{MCHB2DLDPHI1}) boils down to  
a system of {\em non-linear} coupled differential equations. 
Furthermore, in this paper we work in $\frac{\hbar^2}{L^2m}=1$ units, where $m$ is the mass of boson and 
$L$ is the length scale. 

We recall that for MCHB(2) theory the number of particles, the inter-particle interaction strength $\lambda_0$ 
and the external trap potential have to be specified.  
The expansion coefficients $\{C_{(n_1,n_2)}\}$ and orbitals $\phi_1$ and $\phi_2$
are treated as variational parameters.
The optimal values of these parameters are determined {\em self-consistently} 
by diagonalizing the secular Hamiltonian matrix to find the expansion coefficients, Eq.(\ref{MCHB2MATR}),
and by solving the coupled system of non-eigenvalue non-linear differential equations Eq.(\ref{MCHB2DLDPHI1})
to get the one-particle functions.

The self-consistent procedure of finding the optimal one-particle functions 
and corresponding expansion coefficients can be implemented as follows.
We start from some initial guess for the one-particle functions $\phi_1,\phi_2$ obtained, 
say, as the two eigenfunctions lowest in energy of the {\it bare} Hamiltonian 
\begin{equation}
\hat{h}\phi_i=\left [-\frac{1}{2}\nabla^2_{\vec{r}}+V(\vec{r})\right ]\phi_i=\epsilon_i \phi_i,
\label{BASIS1}
\end{equation}
where $i=1,2$ and $V(\vec{r})$ is the corresponding trap potential.
These one-particle functions are used to construct the permanents in the expansion Eq.(\ref{MCHB2PSI})
and to evaluate the Hamiltonian matrix elements.
By diagonalizing this matrix, we solve the corresponding secular equation Eq.(\ref{MCHB2MATR}) and 
get N+1 orthonormal eigenfunctions and the corresponding eigenvalues. 
The lowest-energy solution corresponds to the ground state of the problem, while
other solutions may be used to attack excited states. 
The eigenvector of interest contains the set of expansion coefficients $\{C_{(n_1,n_2)}\}$.
To implement the second part of the variational principle, we use 
these expansion coefficients to evaluate the elements of the reduced one- and two-body densities
given in Eqs.(\ref{DNS1ij},\ref{DNS2ii}) and in Appendix \ref{A1},
required by the coupled system (\ref{MCHB2DLDPHI1}) of the MCHB(2) equations.
By solving this system of equations we find a new set of the one-particle functions $\phi_1,\phi_2$. 
This self-consistent scheme is iterated until convergence is achieved.

One goal here is to demonstrate that 
when the bosonic system is treated at the many-body level and the 
many-body basis set is incomplete, then the choice of one-particle functions used to construct 
this many-body basis set has a great impact on the results and predictions obtained.
To realize this in practice we consider two sets of one-particle functions 
and compare the many-body predictions obtained within each set.
As the first set we use the lowest-energy eigenfunctions of the bare Hamiltonian 
Eq.(\ref{BASIS1}). We stress that such a fixed choice of the one-particle functions 
is often used \cite{Ueda,Schweds,FixOrb} in many-body treatments on bosonic systems.
The MCHB(2) solution is the second set of orbitals which is, of course, the best possible choice 
because these orbitals have been obtained in the framework of the full variational principle.
From now on we use BH and MCHB superscripts to distinguish the results 
obtained within bare Hamiltonian and MCHB(2) one-particle function sets, respectively.
For example, we denote the corresponding one-particle functions as ${\phi^{BH}_1,\phi^{BH}_2}$ and
${\phi^{MCHB}_1,\phi^{MCHB}_2}$.
In our iterative MCHB(2) scheme we use ${\phi^{BH}_1,\phi^{BH}_2}$ as the initial guess.
Therefore, by comparing the results obtained at the first and last iterations 
we immediately observe the impact of self-consistency.

Let us now elaborate on the choice of trapping potentials.
In our work we have chosen the following form of symmetric and asymmetric double-well potentials.
Let us imagine two separate trap potentials, for simplicity in 1D, described by $V_1(x)$ and $V_2(x)$.
We construct a "superposition" of these traps in such a way that, on the one hand,
the profiles of each potential well in the resulting double-well potential
would be as much as possible close to the original potentials. On the other hand,
it is also desirable that the inter-trap separation, degree of asymmetry 
and barrier height can be easily manipulated.
What is also important is that this double-well potential would have a simple (differentiable) analytic form
and permit 2D and 3D generalizations.
Let us diagonalize the matrix
\begin{equation}
 \left( \begin{array}{cc}
V_1(x+x_0)+B & C  \\
C & V_2(x-x_0)  \end{array} \right).
\label{STrapG}
\end{equation}
The lowest eigenvalue of this matrix is a function which fulfils almost all the above mentioned requirements.
The original traps are connected to each other and the parameter $C$ is responsible for the smoothness
of this connection. The minima of the wells are located approximatively at $\pm x_0$ and
the $V_1(x)$ trap is biased by $B$ with respect to the $V_2(x)$ one.
However, to gain an additional control on the barrier height we add a smooth barrier function 
centered at the origin: 
\begin{equation}
V_{b}(A,D)=\frac{A}{\sqrt{2 \pi} D}\exp{\frac{-x^2}{2 D^2}},
\label{GB}
\end{equation}
where $D$ defines the width of this additional barrier function and $A$ can be used to vary its height.
The resulting trap potential takes on an analytic form:
$$
V(x)= \frac{1}{2}[V_1(x+x_0)+B- \sqrt{4C^2+(V_1(x+x_0)+B-V_2(x-x_0))^2}+V_2(x-x_0)]+ V_{b}(A,D).
$$
We recall that in this paper we work in $\frac{\hbar^2}{L^2m}=1$ units, where $m$ is the mass of boson and 
$L$ is the length scale.

\subsection{Ground state in 1D symmetric double-well trap} \label{SecIV.II}

Let us first apply MCHB(2) theory to study the ground state 
of repulsive bosons trapped in a symmetric double-well potential.
This is a popular problem, widely discussed in the literature, see Refs.\cite{SpeSip,Reinhard,DW,DWQ} and
references therein. We consider a system of N=1000 bosons which is of the order 
of the particle number taken in recent experiments \cite{Design1}.
This system is trapped in the symmetric double-well potential:
\begin{equation}
V_{symm}(x)=0.5x^2-\frac{1}{2}\sqrt{25+64x^2}+8.19531+V_b(A=4,D=0.75).
\label{Straps1D}
\end{equation}
This potential plotted in the left panel of Fig.\ref{fig1}
by a thick solid(black) line is obtained by 
diagonalizing Eq.(\ref{STrapG}) with $V_1(x)=0.5(x+4.0)^2$, $V_2(x)=0.5(x-4.0)^2$, $C=2.5$, $B=0.0$ 
and by adding a barrier function $V_b(A=4,D=0.75)$ (see Eq.(\ref{GB})).
A constant shift has been applied to put the minimum of the potential energy to zero.

In the right panel of Fig.\ref{fig1} we plot the ground state energies per particle
of this system as a function of the inter-particle interaction strength $\lambda_0$.
The many-body results obtained within fixed bare Hamiltonian (BH) one-particle functions
are plotted by a solid (black) line with open circles. The solid (red) line with filled circles
represents the energies per particle obtained self-consistently within the framework of the MCHB(2) theory. 
Both curves coincide only at the limit of non-interacting particles,
the optimal one-particle functions in this case are the bare Hamiltonian functions. Increasing $\lambda_0$, 
the MCHB(2) energy curve gradually (exponentially) deviates from the fixed-orbital one.
To illustrate this we plot in the inset 
the difference between these energies per particle $\Delta=E^{BH}-E^{MCHB}$ 
as a function of $\lambda_0$ (notice the logarithmic scale on both axes).
It is clearly seen that the many-body treatment with fixed one-particle functions 
provides an adequate description only up to $\lambda_0\sim10^{-3}$. For stronger inter-particle
interactions the self-consistency becomes more and more relevant, and leads to lower energies.
We conclude that the choice of the one particle functions 
in truncated CI expansions 
is very crucial for the appropriate description of the energetics of interacting bosonic systems.

Next, we construct the reduced one-particle density matrix 
Eq.(\ref{DNS1ij}) and find the corresponding natural orbitals and natural occupation numbers Eq.(\ref{NOA}).
The many-body ansatz used in the MCHB(2) theory implies that there are only two natural orbitals
with respective occupations numbers $\rho_1$ and $\rho_2$. The conservation of the total number of particles 
$\rho_1+\rho_2=N$ allows us to consider only one occupation number.

In Fig.\ref{fig2} we plot the occupation number $\rho_2$ of the second natural orbital
as a function of the inter-particle interaction strength $\lambda_0$.
The solid (red) line with filled circles and the solid (black) line with open circles
represent the self-consistent MCHB(2) and fixed-orbital BH results, respectively.
Up to $\lambda_0\sim10^{-3}$ both methods predict that the value $\rho_2$ gradually increases with $\lambda_0$.
In other words, according to both many-body treatments fragmentation 
takes place when $\lambda_0$ is increased and at $\lambda_0\sim10^{-3}$ approximately 200 
particles out of $N=1000$ are fragmented.
However, for $\lambda_0\ge10^{-3}$ the predictions obtained within the self-consistent MCHB(2) and 
the fixed-orbital many-body theory start to deviate from each other and eventually
{\it drastically contradict} each other.
The many-body results obtained with fixed bare Hamiltonian functions show
that the fragmented fraction $\rho_2$ increases further with $\lambda_0$ until it saturates to some constant value,
$\rho_2\approx366$.
In contrast, the MCHB(2) theory predicts that at some inter-particle interaction strength 
the $\rho_2$ fragmented fraction approaches its maximum value and then gradually decreases.
Finally, we would like to mention that at much larger values of $\lambda_0$ another physical 
phenomenon starts to take place, fermionization \cite{Path, Fermionization},
but the region of inter-particle interaction strengths studied here is far below this limit.

Now we elaborate on the physics behind these many-body predictions.
The ground state fragmentation phenomenon studied here appears due 
to the double-well topology of the trap potential and disappears at zero barrier height.
With increasing inter-particle interaction the respective chemical potential(s) 
of the trapped repulsive bosons increase as well and this 
can be viewed as an effective reduction of the barrier height. 
The trap potential used in the present study has a barrier of finite height and, hence, from 
some critical interaction strength on the bosons are energetically above the barrier and
do not "see" it.  
We may thus conclude that the self-consistent MCHB(2) theory 
predicts a physically relevant 
behavior of the fragmentation as a function of inter-particle interaction strength
in contrast to that predicted by the fixed bare Hamiltonian functions.

Now we investigate the fragmentation phenomenon in the symmetric double-well potential 
as a function of the barrier height. We ask the question at which barrier height 
the system of $N=1000$ bosons at fixed inter-particle interaction strength of
$\lambda_0=0.01$ becomes, say, $25\%$ fragmented.
We consider the same symmetric double-well trap potential as in Eq.(\ref{Straps1D}) and 
rump the barrier up by varying the parameter $A$ of $V_b(A,D=0.75)$ (see Eq.(\ref{GB})). 
For every $A$ a constant shift is applied to put the minimum of the respective potential energy to zero, 
hence $V_{symm}(x)$ at $x=0$ determines the value of the barrier height. 
In Fig.\ref{fig3} we plot the occupation number $\rho_2$ of the second natural orbital
as a function of the barrier height.
The solid (red) line with filled circles and the solid (black) line with open circles 
represent the MCHB(2) and fixed-orbital BH results, respectively.
In this figure it is clearly seen that $25\%$ fragmentation ratio, i.e., $\rho_2=250$ out of $N=1000$ bosons,
is obtained in the framework of fixed-orbital many-body theory at a barrier height $V_{symm}(0)\approx6.75$,
while the self-consistent MCHB(2) gives such a fragmentation ratio 
when the barrier height is $V_{symm}(0)\approx10.3$.
Again, the BH predictions considerably overestimate the fragmentation.
In reality the fragmentation develops slower
with increasing barrier height than predicted by the fixed-orbital BH many-body method. 
This observation is also of a relevance 
for multi-well systems, including optical lattices.   
We stress that the difference between predictions of both theories has a non-trivial origin,
and one curve can not be obtain from the other by a simple procedure, e.g. shift.
To illustrate this, we plot in Fig.\ref{fig3} by dashed (blue) line the difference between 
corresponding occupation numbers $\Delta=\rho_2^{BH}-\rho_2^{MCHB}$ as a function of the barrier height.
In this figure we see that this difference is substantial
at any finite barrier heights and becomes less pronounced only in the limit of very large barrier heights. 

In these investigations of the ground state of a bosonic system trapped in symmetric double-wells
we have seen that the predictions obtained within the framework of the fully self-consistent MCHB(2) and 
within fixed-orbital many-body theories utilizing the same active space 
are quantitatively and some times even qualitatively different.
By construction,
the self-consistent description is more precise than the fixed-orbital one. 
The fixed-orbital many-body theory can, in principle, reproduce the self-consistent MCHB(2) results 
if more BH orbitals are included, i.e., if the active space is enlarged, resulting in a
substantial increase of the computational effort which can be in practice beyond reach.

\subsection{Ground state in 1D asymmetric double-well trap}{\label{SecIV.III}

Despite considerable progress in the experimental studies on double-well traps \cite{Design1,Design2,Design3}, 
a realization of a double-well potential with perfect symmetry is not straightforward.
In contrast, bosonic systems trapped in a perfect double-well potential is 
the most popular theoretical problem addressed in the literature \cite{SpeSip,Reinhard,DW,DWQ},
while theoretical studies on bosonic systems in asymmetric traps remain scarce \cite{DWasymm}.
To elaborate on this very complicated problem, 
we address here the ground state of bosonic systems trapped in one-dimensional asymmetric 
double-well trap potentials. 
Again, the goal is to demonstrate that self-consistent many-body methods remain of importance.

To construct the asymmetric double-well trap 
we locate two equal harmonic traps at $\pm x_0$ and displace the left trap upwards with a bias $B$.
Clearly, if the wells are well separated, then the bare eigenfunctions 
of this trap lowest in energy are 
predominantly localized either in the left or right wells and keep the shapes of the pure harmonic functions.
For a comparatively small bias $B$, the three eigenstates of the double-well potential 
lowest in energy are ordered as depicted in the left panel of Fig.\ref{fig5}. 
Such an asymmetric double-well potential can be obtained by diagonalizing Eq.(\ref{STrapG}) 
with $V_1(x)=0.5(x+4.0)^2$, $V_2(x)=0.5(x-4.0)^2$, $C=2.5$, $B=0.5$ 
and by adding a barrier function $V_b(A=4,D=0.75)$.
A constant shift is also applied to put the global minimum of the potential energy to zero.
The analytical function of this potential reads
\begin{equation}
V_{asymm}(x)=0.5x^2-\frac{1}{2}\sqrt{25.25+8x+64x^2}+V_b(A=4,D=0.75)+8.4423.
\label{Asymm1D}
\end{equation}

We consider N=1000 bosons trapped in the asymmetric double-well potential of Eq.(\ref{Asymm1D}).
Let us see how the ground state of this system develops with increasing 
inter-particle interaction strength $\lambda_0$.
The physical picture of this development, also supported by mean-field studies \cite{BMFIMPL,Path} is as follows.
At zero interaction all bosons are localized 
in the deeper (right) well. The bosons continue to stay localized in this well up to some critical 
interaction strength $\lambda^{cr}_0$. From this $\lambda^{cr}_0$ on the tunnelling 
of bosons into the left well becomes possible and bosons start to occupy the left well.
In other words, there are two regimes of $\lambda_0$, in the first regime $\lambda_0<\lambda^{cr}_0$
the ground state properties depend mainly on the geometry of the deeper well, while in the second regime 
$\lambda_0>\lambda^{cr}_0$ they depend on the global geometry of the asymmetric double-well potential.
These observations are supported by our MCHB(2) calculations 
presented in the right panel of Fig.\ref{fig5}. The MCHB(2) theory gives $\lambda^{cr}_0=0.00136$
for the transition between the two regimes.

In the right lower panel of Fig.\ref{fig5} one can see that
in the first regime ($\lambda_0=0.00135<\lambda^{cr}_0$)
both the orbitals and the density are indeed localized in the right well.
The natural analysis tells that $\rho_1=999.98$ bosons are condensed on the 
first natural orbital (red) and the fraction of $\rho_2=0.02$ bosons is depleted  
to the second natural orbital (blue). Since both orbitals are localized in the same well, 
we can say that the origin of the depletion is on-site excitations.
The MCHB natural orbitals and the density for the second regime ($\lambda_0=0.01>\lambda^{cr}_0$)
are plotted in the upper right panel of Fig.\ref{fig5} and as it was expected in this case both wells are populated.
This ground state is almost $10\%$ fragmented, because $\rho_1=906.06$ bosons 
reside in the first and $\rho_2=93.94$ bosons in the second natural orbital. 

Let us see whether it is possible at all to obtain qualitatively 
the same results by using the fixed-orbital many-body method.
The bare Hamiltonian functions of the asymmetric double-well potential 
of Eq.(\ref{Asymm1D}) are depicted schematically in the left panel of Fig.\ref{fig5}.
If the first and second orbitals are used to construct the permanents, 
then at any non-zero interaction strength
the bosons are spread over both wells. Consequently, with such a choice of
one-particle functions the first regime can not be described. 
Instead, one can try to use the first and third eigenfunctions 
of the bare Hamiltonian to construct the many-body basis set. 
In this case, however, it is impossible to address the second regime. 
To overcome this difficulty one can use all three orbitals simultaneously.
However the active space, i.e., the number of many-body basis functions used
in this case is much larger then the MCHB(2) ones. For $N=1000$ we would need 
$\left( \begin{array}{c} 3+1000-1 \\ 1000\\ \end{array} \right)=501501$ configurations instead of 1001!
Still, the self-consistency is not used and the quality of these fixed-orbital results has to be investigated.  

\subsection{Distributions of the expansion coefficients}{\label{SecIV.IV}

So far, to study bosonic systems trapped in the symmetric and asymmetric
double-well potentials we have considered and analyzed the MCHB energies and orbitals.
We recall that the MCHB solution is given by
the optimal sets of one-particle functions and of the respective expansion coefficients
obtained self-consistently.
In this section we analyze in some detail the MCHB coefficients $\{C_{(n_1,\cdots,n_M)}\}$ 
appearing in the expansion Eq.(\ref{PSIexpansII}) and 
exploit the freedom of unitary transformation as put forward in Sec.\ref{SecIII.C}. 

Generally, for any orthogonal many-body basis set the square of the expansion coefficient
$C^*_{(n_1,\cdots,n_M)}C_{(n_1,\cdots,n_M)}$ defines the probability to find the many-body solution
in the configuration described by the respective many-body basis function $|n_1,n_2,n_3,\cdots,n_M\rangle$.
In other words, the squares of the expansion coefficients can be considered
as a multidimensional discrete probability distribution in the
discrete sample space spanned by the many-body basis functions.
In the  MCHB(2) theory, to count all many-body basis functions
$|n_1,n_2\rangle\equiv|n_1,N-n_1\rangle$ one needs only one independent parameter $n_1$ ($n_2=N-n_1$).
Therefore, the squares of expansion coefficients can be viewed as
a probability function $P(n)$ of a discrete distribution defined over 
the independent parameter $n=0,1,2,\cdots,N$:
\begin{equation}
P(n)=C^*_{(n,N-n)}C_{(n,N-n)}.
\label{D}
\end{equation}
We can use the mean value $\nu$ and variance $\sigma^2$ as measures of the distribution:
\begin{eqnarray}
\nu_{n_1}&=&\sum_{n_1=0}^{N}P(n_1)n_1=\sum_{n_1=0}^{N}C_{(n_1,n_2)}^*C_{(n_1,n_2)} n_1
\equiv\langle \hat n_1 \rangle, \nonumber\\
\sigma^2_{n_1}&=&\sum_{n_1=0}^{N}P(n_1)n_1^2-(\sum_{n_1=0}^{N}P(n_1)n_1)^2
\equiv\langle \hat n_1^2 \rangle-\langle \hat n_1 \rangle^2.
\label{MeanVariance}
\end{eqnarray}
Here, we use $n_1$ as the independent parameter. 
If the occupation number of the second orbital $n_2$ is used instead then $\nu_{n_2}=N-\nu_{n_1}$.
Interestingly, the $\langle \hat n_i \rangle$  and $\langle \hat n_i^2 \rangle$
have already appeared in the evaluation of the diagonal elements of the
reduced one-particle Eq.(\ref{DNS1ij}) and two-particle Eq.(\ref{DNS2ii}) density matrices.

As mentioned above in Sec.\ref{SecIII.C},
due to the invariance of the MCHB solution with respect to unitary transformations,
there are infinitely many possible choices of the MCHB orbitals which give the same energy.
It is also clear that the distribution of the expansion coefficients depends on
the particular choice of the one-particle functions used to construct the many-body basis set.
Therefore, there are infinitely many possible distributions of the expansion coefficients as well.
However, since the mean values and variances of the distributions are different,
we use these quantities as the main criteria to compare energetically equivalent distributions. 
The main aim now is to find the distributions characterized by the {\em minimal width}.

Let us consider two sets of the MCHB(2) orbitals connected by a unitary (orthogonal) transformation
\begin{equation}
\left(
\begin{array}{c}
\tilde\phi_1 \\
\tilde\phi_2
\end{array}
\right)
=\cal\hat{U}
\left(
\begin{array}{c}
\phi_1 \\
\phi_2
\end{array}
\right)
\equiv
\left(
\begin{matrix}
\cos{\theta} & \sin{\theta} \cr -\sin{\theta} &  \cos{\theta}
\end{matrix}
\right)
\left(
\begin{array}{c}
\phi_1 \\
\phi_2
\end{array}
\right),
\label{optangle}
\end{equation}
where $\theta$ is the rotation angle.
Clearly, the creation and annihilation operators corresponding to each set of the orbitals
are coupled via $\cal\hat{U}$ as well:
\begin{equation}
\left(
\begin{array}{c}
\tilde{b}^{(\dagger)}_1 \\
\tilde{b}^{(\dagger)}_2
\end{array}
\right)
=
\left(
\begin{matrix}
\cos{\theta} & \sin{\theta} \cr -\sin{\theta} &  \cos{\theta}
\end{matrix}
\right)
\left(
\begin{array}{c}
b^{(\dagger)}_1 \\
b^{(\dagger)}_2
\end{array}
\right).
\label{optoper}
\end{equation}
Since all possible real MCHB(2) orbitals can be obtained from the initial $(\phi_1,\phi_2)$ ones
by changing the angle $\theta$, the respective distributions of the expansion coefficients
as well as their mean values and variances depend also on this angle.
Here we are interested in the variance:
\begin{eqnarray}
\tilde\sigma^2_{n_1}&\equiv& \langle \tilde{\hat n}_1^2 \rangle-\langle \tilde{\hat n}_1 \rangle^2=
\langle \tilde{b}_1^\dagger \tilde{b}_1 \tilde{b}_1^\dagger
\tilde{b}_1 \rangle-\langle \tilde{b}_1^\dagger \tilde{b}_1 \rangle^2=\sigma^2(\theta).
\label{VarianceTheta}
\end{eqnarray}
After straightforward algebra one finds
\begin{eqnarray}
\sigma^2(\theta)&=& \cos^4\theta (\langle \hat n_1^2 \rangle - \langle \hat n_1 \rangle^2 )
+\sin^4\theta(\langle \hat n_2^2\rangle-\langle \hat n_2\rangle^2) \nonumber\\
&+&\sin^2\theta\cos^2\theta[4\langle \hat n_1 \hat n_2\rangle-2\langle \hat n_1\rangle \langle \hat n_2\rangle+
N+\rho_{1122}+ \rho_{2211}-(\rho_{12}+\rho_{21})^2] \nonumber\\
 &+& 2 \sin\theta\cos^3\theta[\rho_{12}+\rho_{1112}+\rho_{2111}-(\rho_{12}+\rho_{21})\langle \hat n_1\rangle] 
\nonumber\\
 &+& 2 \sin^3\theta\cos\theta[\rho_{21}+\rho_{1222}+\rho_{2221}-(\rho_{12}+\rho_{21})\langle \hat n_2\rangle], 
\label{VarianceThetaT}
\end{eqnarray}
where the involved diagonal and off-diagonal elements of the 
reduced one- and two-body density matrices are given in 
Eqs.(\ref{DNS1ij},\ref{DNS2ii}) and in Appendix \ref{A1}.
The extrema of this function are obtained in the ordinary way, by solving
\begin{equation}
\frac{\partial}{\partial \theta} \sigma^2(\theta)=0.
\label{minsigma}
\end{equation}

The procedure of finding the distribution with minimal variance is implemented as follows.
Having at hand a MCHB(2) solution, i.e., the orbitals $(\phi_1,\phi_2)$ and
a set of the respective expansion coefficients $\{C_{(n_1,n_2)}\}$,
we recompute all required elements of the reduced one- and two-body density matrices,
appearing in Eq.(\ref{VarianceThetaT}) and
explicitly reconstruct the variance $\sigma^2(\theta)$.
The angle $\theta_{min}$ at which this function has a minimum is obtained
numerically by solving Eq.(\ref{minsigma}).
The unitary transformation Eq.(\ref{optangle}) with this angle
gives a new set of MCHB(2) orbitals $(\tilde\phi_1,\tilde\phi_2)$.
The distribution of expansion coefficients computed on these orbitals
has the minimal variance $\sigma_{min}$. Obviously, we can call such a distribution the {\em minimal}
distribution.

The natural orbitals being the MCHB solutions can be used 
to shed light on the physical nature (depletion or fragmentation) of the ground and excited states.
Let us see how the expansion coefficients distributions 
of the many-body basis sets constructed by using the natural orbitals look like.
In the left lower panel of Fig.\ref{fig6} we plot the expansion coefficients 
obtained for the ground state of N=1000 bosons with $\lambda_0=0.01$ trapped 
in the symmetric double-well potential Eq.(\ref{Straps1D}).
This system has been discussed above in Sec.\ref{SecIV.II}. 
The corresponding ground state natural orbitals are very similar to those plotted 
in the right lower panel of Fig.\ref{fig4}. 
Due to the symmetry of the trap potential, the ground many-body state is, of course, of gerade symmetry,
while the natural orbitals are of gerade and ungerade symmetries.
Therefore, for the ground state the non-zero expansion coefficients can appear 
only due to the contributions of configurations of gerade symmetry. 
Indeed, in the left lower panel of Fig.\ref{fig6} one can see that the configurations  
with even number of bosons residing in the ungerade orbital contribute, while those with odd numbers 
i.e., $|1000-1,1\rangle$, $|1000-3,3\rangle$,$|1000-5,5\rangle$ etc., give zero contributions.  
In this figure it is clearly seen that the main contributions come from  
the first configurations, where almost all bosons reside in the first orbital,
while the configurations with large populations of the second orbital do not contribute.
This observation is supported by
the statistical description of this distribution in terms of its mean values (see Eq.(\ref{MeanVariance})).
The mean statistical values of this distribution are 
$\nu_1=\langle \hat n_1 \rangle=994.78$ and $\nu_2=\langle \hat n_2 \rangle=5.22$.
As one would expect these mean values are identical to the natural occupation 
numbers of the respective natural orbitals.

In the right lower panel of Fig.\ref{fig6} we plot the expansion coefficients
obtained for the system of N=1000 bosons with $\lambda_0=0.01$ trapped 
in the asymmetric double-well potential Eq.(\ref{Asymm1D}) 
discussed in Sec.\ref{SecIV.III}. The corresponding natural orbitals
presented in the right upper panel of Fig.\ref{fig5} do not possess any symmetry.
Consequently, all the many-body basis functions constructed by using these natural orbitals
can give non-zero contributions in the many-body expansion. Indeed, we can see this in the right lower 
panel of Fig.\ref{fig6}. The mean statistical values of this asymmetric distribution 
$\nu_1=\langle \hat n_1 \rangle=906.06$ and $\nu_2=\langle \hat n_2 \rangle=93.94$ 
are, of course,
identical to the respective natural orbital occupation numbers. 

Let us see how the variance $\sigma^2_{n_1}=\langle \hat n_1^2 \rangle-\langle \hat n_1 \rangle^2$
characterizes the distributions of the expansion coefficients.
Inspecting the distributions obtained with natural orbitals 
we see in the lower panels of Fig.\ref{fig6} that for the symmetric double-well, where 
only several many-body basis function have non-zero expansion coefficients,
the variance is quite small $\sigma_{NO}=7.654$. While 
in the asymmetric case, where almost all expansion coefficients have non-zero contributions,
the width of the distribution is much larger $\sigma_{NO}=121.731$. We conclude that the value
of the statistical variance $\sigma^2$, characterizing the width of the distribution of the
expansion coefficients, 
indeed provides an adequate estimation on the number of contributing configurations.

Now let us see how the minimal distributions, i.e., the distributions with minimal width look like.
We consider the same examples of the symmetric and asymmetric double-wells as above,
and find their minimal distributions by using the developed algorithm (see Eq.(\ref{minsigma})).
The minimal distributions obtained for the symmetric and asymmetric 
double-wells are presented in the left and right upper panels of Fig.\ref{fig6} respectively.
The respective variances are also depicted.
The minimal distribution obtained for the symmetric double-well trap has a maximum
located exactly at the $|500,500 \rangle$.
This is a symmetric distribution because the pairs of basis vectors around maximum 
$|500-i,500+i \rangle$ and $|500+i,500-i \rangle$ contribute with identical coefficients. 
Interestingly, the minimal distribution is smooth, in contrast to 
that obtained within the natural orbitals, where the neighboring MCHB(2) coefficients 
are of different sign, see lower left panel of Fig.\ref{fig6}. 
The width of the minimal distribution is, of course, smaller than that for the
natural orbital, $\sigma_{min}=3.313$ in comparison to $\sigma_{NO}=7.654$.

Comparing the right upper and lower panels of Fig.\ref{fig6} we see that 
in the presented example of asymmetric double-well trap the minimal distribution
differs substantially from that obtained within the natural orbitals.
For the asymmetric case, the width of the minimal distribution $\sigma_{min}=0.753$ 
is about $2.5$ orders of magnitude smaller than that for the natural orbital $\sigma_{NO}=121.731$.
This means that the minimal distribution is formed by several configurations
in contrast to the broad distribution obtained with natural orbitals where all configurations contribute.
The maximum of the minimal distribution is located around $|600,400 \rangle$.
The distribution is smooth and, of course, not symmetric.

Irrespective of the symmetry of the trap potential used,
the width of the minimal distribution can be much smaller than that obtained with the natural orbitals.
The profile of the minimal distribution is smooth, while the distribution 
of the expansion coefficients obtained within other orbital sets is not necessarily so.
These observations lead to several important consequences.
First, the minimal distribution of the expansion coefficients can be approximated by a smooth continuous 
function. 
Looking at the pictures in Fig.\ref{fig6} we approximate the probability function Eq.(\ref{D}) 
of the minimal distribution by a Gaussian function: 
\begin{eqnarray}
P(\xi)=\frac{1}{\sqrt{2 \pi}\sigma}\exp{\frac{-(\xi-\xi_0)^2}{2\sigma^2}}.
\label{GaussApprox} 
\end{eqnarray}
The parameters of this function are obtained straightforwardly.
We take $\xi=n_1$ as an independent variable. The averages Eq.(\ref{MeanVariance}) of the minimal distribution
define the location $\xi_0=\langle \hat n_1\rangle$ and width $\sigma=\sigma_{min}$ of the Gaussian.
In the upper panels of Fig.\ref{fig6} we plot the Gaussian distribution functions approximating  
the minimal distributions of the studied symmetric and asymmetric systems by black solid lines.
We stress that in Fig.\ref{fig6} we plot the distributions of the expansion coefficients, i.e.,
$\sqrt{P(n_1)}$. 
In this figure we see that the Gaussian distributions match the numerical results very well.
Moreover, once a MCHB calculation has been performed, 
this continuous Gaussian approximation does not require any fitting parameters.

We mention that continuous approximations to the discrete 
distributions of the expansion coefficients have already been addressed in the literature 
for symmetric trap potentials \cite{SpeSip,ConApp}. In these studies 
{\it smoothness} of the real CI coefficients and thus of the continuum distribution 
approximation was used as a basic {\it assumption}.
In contrast, here we have demonstrated numerically that the profile of 
the distribution of the expansion coefficients obtained within MCHB many-body method can indeed be smooth.
Moreover, the smooth profiles of the discrete distribution of the expansion coefficients 
are observed in examples of symmetric as well as of asymmetric trap potentials.
However, it is very important to stress that smoothness  
is achieved only for the minimal distributions, i.e.,
for a very specific choice of the orbitals (see Eq.(\ref{VarianceThetaT})).

Finally, we remark that the existence of smooth continuous functions
approximating the discrete distribution of the expansion coefficients 
makes the developments of self-consistent many-body methods very promising
for attacking many-particle bosonic systems within huge configurational spaces.

\subsection{Excited states in 1D symmetric double-well trap }{\label{SecIV.V}

In the preceding sections we considered the ground state of a trapped bosonic system and
addressed condensation and fragmentation as properties of the ground state.
The main subject of this part of our work is to touch upon properties of excited states of trapped bosons.
The studies on excited states 
are of great interest \cite{ModHam1,ModHam2,ModHam3,ModHam4,Bragg} because of their relevance for 
depletion and stability of condensates, for time-dependent and finite-temperature effects,
for formations of solitons and soliton trains \cite{Solitons}, 
as well as for other interesting phenomena.

Let us assume that we have obtained the self-consistent MCHB orbitals for the ground state of the bosonic system.
Since in the MCHB scheme the diagonalization of the secular matrix was employed we also 
have the energies and many-body wave-functions of the excited states. 
However, the excited states computed in this way are {\it not} self-consistent. 
Here we address the question whether the self-consistent orbitals obtained for the ground state
can also be used to provide an adequate description of the excited states or whether
every excited state has to be treated individually.

To answer this question we consider a system of N=1000 bosons with 
$\lambda_0=0.01$ trapped in a symmetric double-well potential.
In this example we use the trap:
$$
V_{symm}(x)=0.5x^2-\frac{1}{2}\sqrt{25+64x^2}+V_b(A=25,D=0.75)+8.19523,
\nonumber
$$
obtained by diagonalizing Eq.(\ref{STrapG}) with $V_1(x)=0.5(x+4.0)^2$, $V_2(x)=0.5(x-4.0)^2$, $C=2.5$, $B=0.0$ 
and by adding a barrier function $V_b(A=25,D=0.75)$.
To put the minimum of the potential energy to zero we also use a constant shift.

First, we apply the MCHB(2) approach to obtain the self-consistent energy and orbitals of the ground state.
This state is essentially condensed because the occupation numbers of the corresponding 
reduced one-particle density matrix are $\rho_1=994.78$ and $\rho_2=5.22$. 
In other words, 994.78 particles are condensed in the
first natural orbital and 5.22 are excited out to the second orbital.
The density and respective natural orbitals are plotted in the lower right panel of Fig.\ref{fig4}.
We recall that the natural orbitals are solutions of the MCHB.

Having at hand the self-consistent ground state orbitals we diagonalize the 
full secular matrix and get the energies of the excited states.  
In the left panel of Fig.\ref{fig4} we depict the energy level of the first excited state. 
We connect both levels by a vertical solid line with arrow to stress that this first excited state 
is obtained by using the MCHB orbitals of the ground state.
The natural orbital analysis applied reveals that 
the occupations numbers of this first excited state are $\rho_1=985.20$, $\rho_2=14.80$ and
the natural orbitals of this excited state are almost identical to the ground state ones.
Comparing the natural orbital occupation numbers of the 
ground and this first excited state we conclude that 
this excited state can be considered as a further microscopic excitation of 
a small number of particles out of the condensate, i.e., a further depletion of the condensate.

Let us now see what happens when the first excited state is treated self-consistently.
To realize this we employ the developed MCHB(2) method to the excited states as follows. 
We recall that in our iterative MCHB scheme the many-body expansion coefficients corresponding to 
the ground state are obtained as components of the first, i.e., lowest-energy eigenvector of the secular matrix.
To attack the first excited state we use the components of the second eigenvector during 
all iterations.
The self-consistent results obtained by applying this procedure to the first excited state 
are also depicted in Fig.\ref{fig4} and marked as "SC Excited State". 
The natural orbitals and the density corresponding to this state are shown in the upper right panel
and its energy level is depicted in the left panel.

From Fig.\ref{fig4} it is clearly seen that self-consistency 
can have an enormous impact on properties of an excited state.
The energy of the first excited state obtained self-consistently is much lower than that
obtained by using the ground state orbitals.
The self-consistent first excited state is almost $30\%$ fragmented. 
The respective natural orbital occupation numbers 
are $\rho_1=734.84$ and $\rho_2=265.16$, contrasting $\rho_1=985.20$ and $\rho_2=14.80$
obtained above for the non-self-consistent excited state.
By comparing the upper and lower right panels of Fig.\ref{fig4} we see that 
the shapes of the first natural orbitals (red) in both calculations are almost identical,
while those of the second natural orbitals (blue) differ drastically from each other.
Moreover, it is easily seen in the right upper panel of Fig.\ref{fig4} that 
a simple linear combination of the natural orbitals of the first self-consistent excited state 
gives almost pure left and right localized functions, which are, of course, solutions of the MCHB as well.
For the ground state orbitals depicted in the right lower panel of Fig.\ref{fig4} such a localized presentation can
not be obtained.
From this analysis we conclude that the first self-consistent excited state and the ground state
are qualitatively different and exhibit a different "topology".

In this example the ground state of the system is condensed and 
the first excited state is fragmented. Fragmentation is shown to be much more favorable energetically
than a further depletion of the condensate.  
One goal of this study is to demonstrate that 
self-consistency can be very important for an adequate description of excited states.
Indeed, we have shown that without self-consistency the lowest-energy excited state
describes a depletion of the condensate.
The inclusion of self-consistency significantly lowers the energy of the first excited state 
and drastically changes its character. Instead of depletion of the condensate it describes its fragmentation.
We stress that excited states of different topology also exist 
in many other trapped bosonic systems. The question whether they are low-lying or highly excited states 
depends on trap geometries, number of particles and strength of inter-particle interaction.

\subsection{Ground state in 2D symmetric double-well trap}\label{SecIV.VI}

In the present section we investigate the relevance of self-consistency 
for many-body studies on trapped bosonic systems in higher dimensions.
For illustration, here we investigate a repulsive bosonic system 
trapped in the two-dimensional symmetric double-well potential
\begin{equation}
V_{symm}(x,y)=0.5x^2+0.5y^2-\frac{1}{2}\sqrt{25+64x^2}+8.19531+V_b(A=8,D=0.75).
\label{Straps2D}
\end{equation}
This potential is obtained according to Eq.(\ref{STrapG}) as a superposition of two pure harmonic 2D potentials
$V_1(x,y)=0.5[(x+4.0)^2+y^2]$ and $V_2(x,y)=0.5[(x-4.0)^2+y^2]$  
with $C=2.5$, $B=0.0$ and by adding the two-dimensional barrier function
$V_b(A,D)=\frac{4}{\sqrt{2\pi}D}\exp{[-(x+y)^2/(2 D^2)]}$.
Here, we have also applied a constant shift to put the minimum of the potential energy to zero.

The ground state of N=1000 identical bosons at $\lambda_0=0.01$
trapped in this double-well trap has been investigated within the framework of
the fixed-orbital and self-consistent MCHB(2) many-body methods.
The two eigenfunctions lowest in energy of the respective two dimensional bare Hamiltonian Eq.(\ref{BASIS1})
have been used to construct the fixed-orbital many-body basis set.
As in the one-dimensional case, we use these functions as an initial guess for solving MCHB(2) equations.

The geometry of the double-well trap used implies that the ground state density is
made of two parts each localized in one well. Due to the perfect two-fold symmetry of the trap potential
it suffices to consider only one of them without loss of information.
In Fig.\ref{fig2D}, for convenience of comparison, we depict  
the part of the self-consistent MCHB(2) ground state density localized in the left well together with
the part of the total density obtained within the usual fixed-orbital many-body method
localized in the right well.
From this figure it is clearly seen that the densities obtained are very different.
To better account for the repulsion between the bosons,
the self-consistent density is more delocalized within the well than the fixed orbital one.
Of course, the MCHB(2) solution has a lower energy than the fixed orbital one. 
The natural orbital analysis applied shows that in the MCHB(2) case 
$\rho_1=750$ bosons reside in one orbital and $\rho_2=250$ bosons in the other orbital, while
for the many-body solution obtained with fixed bare Hamiltonian orbitals these occupations 
are $\rho_1=634$ and $\rho_2=366$.
Both many-body methods give the same qualitative
prediction on the nature of the ground state -- this state is fragmented,
however, the predicted details of the fragmentation are very different.

In this investigation we have demonstrated that analogously to the 1D case, 
self-consistency is of great relevance for the many-body description of bosonic systems in
higher dimensions.

\section{Implementation of the MCHB(M) for two bosons in a trap}\label{SecV}
The technical realization of the developed MCHB method for the general case of M orbitals and N bosons
requires considerable methodological and algorithmical efforts. In this section 
we perform the first step in this direction and implement the MCHB(M) theory for two interacting bosons.

We consider a system of two bosons interacting via contact inter-particle potential 
trapped in the 1D symmetric double-well potential
\begin{equation}
V_{symm}(x)=0.5x^2-2.5\sqrt{0.0784+x^2}+3.16204+V_b(A=1,D=0.75).
\label{Straps2B}
\end{equation}
To obtain this potential we take the lowest-energy eigenvalue of Eq.(\ref{STrapG}) 
with $V_1(x)=0.5(x+2.5)^2$, $V_2(x)=0.5(x-2.5)^2$, $C=0.7$, $B=0.0$ 
and add a barrier function $V_b(A=1,D=0.75)$.
The minimal value of this potential energy has been adjusted to zero by applying a constant shift. 

In Fig.\ref{fig7} we plot the ground state energy of this system as a function of the inter-particle
interaction strength $\lambda_0$, obtained within the framework of the self-consistent 
many-body MCHB(M) method, M=2,$\cdots$,10. We also present the energies obtained 
within one- and two-orbital mean-fields, MF(1) and MF(2) respectively.
All the energies are plotted with respect to the ground state energy $E(0)$ of the 
non-interacting two-boson system.
At this limit the ground state many-body wave-function is given by a single configuration  
where both bosons reside in the lowest-energy orbital of the respective bare Hamiltonian,
and the total energy of this ground state $E(0)$ is twice the respective orbital energy.

In this figure it is very difficult to distinguish between the energy curves obtained within the two-orbital
MCHB(2) theory and the multi-orbital MCHB(M) M=3,$\cdots$,10 ones.
This observation tells us that for our example an adequate description  
can already be obtained within the two-orbital MCHB(2) method,
the inclusion of more orbitals does not lead to significant "observable" improvements.
We demonstrate this in the inset of Fig.\ref{fig7}, where 
we present a part of the MCHB(M) energy curves on an enlarged scale.
Comparing the energy gaps between successive MCHB(M) energy curves,
we observe the convergence of the MCHB(M) method.

To arrive at a deeper insight into the role of many-body effects we also 
study the trapped two-bosonic system within the framework of multi-orbital mean-field theory which 
is a limiting one-permanent case of the MCHB approach as we have shown in Sec.\ref{SecIII.C}.
We recall that the one-orbital mean-field MF(1) is the famous Gross-Pitaevskii mean-field.
In Fig.\ref{fig7} the MF(1) energy curve is depicted by a dashed line.
The one-orbital mean-field solution describes a "condensate" where both bosons reside in the same orbital.
The two-orbital mean-field MF(2) solution describes a situation where one boson resides in one orbital 
and the second boson occupies another orbital. For the two-boson system such a state  
can be considered as a two-fold "fragmented" state.
Here we observe a "critical phenomenon". Up to some critical value of the inter-particle interaction strength
the "condensed" solution is energetically more favorable than the "fragmented" one. 
From this critical $\lambda_0$ on,
the ground state becomes two-fold "fragmented". The intersection of the MF(1) and MF(2) energy curves
gives the critical interaction strength at $\lambda_0=0.0203$. In Fig.\ref{fig7} we mark this point by a
cross.

The MCHB theory gives the following many-body picture of this transition.
The natural orbital analysis applied to the MCHB(M) solutions at each $\lambda_0$ 
reveals that the character of the many-body MCHB(M) ground state smoothly develops 
with inter-particle interaction strength from "condensed", where only one natural orbital is occupied, 
to the two-fold "fragmented" where two natural orbital have dominant and nearly equal occupations.
Having at hand the natural orbital occupation numbers as a function of $\lambda_0$,
we find the inflection point at $\lambda_0=0.0142$ and attribute it to the transition point.
In Fig.\ref{fig7} this point is marked by a bold filled circle. 
The comparison between many-body and mean-field predictions shows that
in contrast to the sharp transition obtained within the multi-orbital mean-fields,
an inclusion of the many-body effect makes this transition smooth, i.e., it is a crossover.
 
This investigation of the minimal many-body system has verified that
qualitative predictions on the transition from condensation to fragmentation in symmetric double wells 
can already be obtained within the framework of the self-consistent multi-orbital mean-field theory 
\cite{BMF,BMFIMPL,ASCPL2005}.
We also demonstrated that the two-orbital MCHB(2) theory provides 
in this case an accurate quantitative description and the inclusion of more orbitals leads to minor changes.
Generally, MCHB(M) opens the door to treat any two-boson system.

\section{Discussion and conclusions}\label{SecVI}
In this work we developed a complete variational many-body theory for systems of $N$ trapped bosons
interacting via a general two-body interaction potential. The many-body wave function of this system
is expanded over orthogonal many-body basis functions (configurations).  
Each basis function is constructed as a symmetrized Hartree product (permanent) 
with N bosons distributed over M one-particle functions. 
These one-particle functions {\em and} the respective expansion coefficients 
are treated as the variational parameters in this theory. 
The optimal variational parameters are obtained {\em self-consistently} by solving 
a coupled system of non-eigenvalue integro-differential equations to get the one-particle functions and,
by diagonalizing the secular Hamiltonian matrix problem, to find the expansion coefficients.
To construct this matrix we derived {\em general} rules for matrix elements,
which are of relevance also for other many-body theories.
We call this self-consistent theory multi-configurational Hartree for bosons, or MCHB(M)
where M specifies the number of one-particle functions involved.

The properties of the MCHB(M) equations were discussed.
These equations are formally also valid for any size of the many-body expansion, i.e.,
for any number of configurations used in the expansion.
Therefore, the MCHB(M) theory allows to find also the best possible 
many-body solution within any restricted configurational space used.
We have shown that in the limiting case where only one permanent forms the many-body expansion,
the MCHB(M) theory boils down to the self-consistent mean fields.
In the simplest case when all bosons reside in the same orbital one gets the Gross-Pitaevskii equation.
The multi-orbital mean-field theory is obtained in the more general single-permanent case,
when bosons are allowed to reside in several one-particle functions.

We have shown that if the many-body basis set spans a complete subspace of the Hilbert space, 
namely, when all possible configurations appearing as permutations of N bosons over M orbitals
are taken into account, then the MCHB(M) solution is invariant with respect to 
a unitary transformation (linear combination) of the MCHB(M) orbitals.
This property has been used to demonstrate that eigenfunctions of the reduced one-particle 
density, i.e., the natural orbitals are the MCHB(M) solution as well.
We proposed to analyze the ground and excited states 
in the terms of the natural orbitals and natural occupation numbers 
to more easily identify the depletion and fragmentation of the condensates.

In the second part of our work we implemented the MCHB(M) method with M=2 orbitals.
We applied it to study the ground and excited states of the bosonic systems 
with the popular contact inter-particle interaction trapped 
in one- and two-dimensional symmetric and asymmetric double-well traps.
The considered configurational space was spanned by all possible permutations of N=1000 bosons over two orbitals.
Two lowest-energy eigenfunctions of the respective bare Hamiltonian were used to construct 
the often employed fixed-orbital many-body basis set. We compare the fixed-orbital many-body predictions 
with those obtained self-consistently via the MCHB(2) theory to investigate the impact and
relevance of self-consistency.

We performed several ground state studies of the bosonic system trapped in the symmetric double-well trap. 
In the first study, we keep the shape of the symmetric double-well trap potential fixed 
and vary the strength $\lambda_0$ of the inter-particle interaction in order to study
the ground state fragmentation.
We have seen that self-consistent MCHB(2) theory predicts a gradual enhancement of 
the fragmented fraction with $\lambda_0$ up to
some critical inter-particle interaction strength, where the fragmentation achieves its maximal value.
Further increase of $\lambda_0$ causes 
gradual decreasing of the fragmentation,
because the energy of the bosonic system in this regime becomes larger than the potential barrier. 
The many-body result obtained within fixed bare Hamiltonian functions
predicts a gradual enhancement of the fragmentation with increasing $\lambda_0$ followed by 
an unphysical saturation of the fragmented fraction to some constant value. 
In the second study, to investigate the transition point from condensation to 
fragmentation we keep the inter-particle interaction strength fixed and rump 
the barrier up. The critical value of the barrier height 
obtained with bare Hamiltonian functions are considerably 
underestimated in comparison with the more exact, self-consistent, MCHB(2) many-body predictions.
A main conclusion derived from this investigation is that the quantitative characterization of the
ground state properties of the bosonic system trapped in symmetric double-wells
can be obtained in the framework of self-consistent methods,
the fixed-orbital many-body theory utilizing the same size of CI expansion can 
be unreliable even concerning qualitative predictions.

We addressed the ground state properties of the bosonic system trapped in the 
asymmetric double-well potential. In this study we keep the shape of the asymmetric 
double-well trap potential fixed and vary the strength of the inter-particle interaction.
The MCHB(2) theory predicts two regimes for the ground state. In the first one 
the atomic cloud is localized in the deeper well; 
from some critical inter-particle interaction strength on the system enters the second regime where
bosons occupy both wells.
We show that such a picture can not be obtained within a fixed two-orbital many-body treatment.
To overcome this difficulty, one must use at least three fixed orbitals to construct the permanents.
However the active space, i.e., the number of many-body basis functions used
in this case is substantial and often beyond reach.

To exploit the freedom of unitary transformations
we analyzed the distribution of the MCHB(2) expansion coefficients 
obtained for different linear combinations of the MCHB(2) orbitals 
for the ground state of the bosonic systems trapped in the symmetric and asymmetric double-wells.
We verified that statistical means and variances can indeed be used to characterize adequately 
the distributions of the expansion coefficients.
Moreover, we have seen that the distributions with minimal width, obtained by minimizing the variance,
exhibit very smooth profiles, irrespective of the symmetry of the trap potential used.
For the studied examples the profiles of the smooth minimal distributions can very well be approximated 
by continuous Gaussian functions. 

We also investigated the first excited state of the system trapped in the symmetric double-well 
and demonstrate that self-consistency can be very important for an adequate description of excited states.
We used the natural orbitals analysis to classify the ground and excited states.
It was shown that without self-consistency the lowest-energy excited state describes a depletion of the condensate.
The inclusion of self-consistency significantly lowers the energy of the first excited state 
and on top of that drastically changes its character: 
Instead of describing a condensate with a slightly larger depleted fraction, it
describes a fragmented condensate with a substantial degree of almost $30\%$ fragmentation.

As an illustrative example we investigate the ground state
of a two dimensional bosonic system trapped in a symmetric double-well potential.
We show that self-consistency is expected to be of high relevance  
for many-body studies on trapped bosonic systems also in higher dimensions.

Finally, we have shown that the two-orbital MCHB(2) theory can provide 
quite accurate quantitative description 
for the ground state of two-bosonic systems in symmetric double-well traps.
The MCHB(M) has been implemented for two bosons and in an illustrative example 
the inclusion of more orbitals leads only to minor changes.

\appendix
\section{Off-diagonal elements of two-body density matrix}\label{A1}
In this appendix we evaluate the off-diagonal elements of the two-body density matrix
$\rho_{ijkl}=\langle \Psi| b_i^\dagger  b_j^\dagger b_k b_l|\Psi \rangle$.
We use the same shorthand notations as defined in Sec.\ref{SecIII.A} where
only occupation numbers of the involved orbitals are shown.
For instance, the configuration $(;n_i-2;n_j+2;)$ differs from $\vec{n}\equiv(;n_i;n_j;)$ by
excitation of two bosons from $\phi_i$ to $\phi_j$.
In all cases different indices can not have the same value. 
$$
\rho_{iijj}=\sum_{\vec{n}}C^*_{\vec{n}} C_{(;n_i-2;n_j+2;)} \sqrt{n_i(n_i-1)(n_j+1)(n_j+2)},
$$
$$
\rho_{iijk}=
\sum_{\vec{n}}C^*_{\vec{n}} C_{(;n_i-2;n_j+1;n_k+1;)} 
\sqrt{n_i(n_i-1)(n_j+1)(n_k+1)}, 
$$
$$
\rho_{ijkl}=
\sum_{\vec{n}}C^*_{\vec{n}} C_{(;n_i-1;n_j-1;n_k+1;n_l+1;)} 
\sqrt{n_in_j(n_k+1)(n_l+1)}, 
$$

$$
\rho_{ijkk}=
\sum_{\vec{n}}C^*_{\vec{n}} C_{(;n_i-1;n_j-1;n_k+2;)} 
\sqrt{n_in_j(n_k+1)(n_k+2)}, 
$$

$$
\rho_{iiij}=
\sum_{\vec{n}}C^*_{\vec{n}} C_{(;n_i-1;n_j+1;)} 
(n_i-1)\sqrt{n_i(n_j+1)}, 
$$

$$
\rho_{ijjj}=
\sum_{\vec{n}}C^*_{\vec{n}} C_{(;n_i-1;n_j+1;)} 
n_j\sqrt{n_i(n_j+1)}, 
$$

$$
\rho_{ikkj}=
\sum_{\vec{n}}C^*_{\vec{n}} C_{(;n_i-1;n_j+1;)} 
n_k\sqrt{n_i(n_j+1)}. 
$$
All other matrix elements are zero or can be reduced to the above ones
due to symmetries of the two-body operator: 
$$
\rho_{ijkl}=\rho_{jikl}=\rho_{ijlk}=\rho_{jilk}.
$$
When the many-body function $|\Psi\rangle$ and one-particle orbitals are real functions,
some additional symmetries are implied:
$
\rho_{12}=\rho_{21},
$
$
\rho_{2111}=\rho_{1112}
$
$
\rho_{2221}=\rho_{1222}
 $
and
$
\rho_{2211}=\rho_{1122}.
$

\begin{acknowledgments}
We acknowledge stimulating discussions with J.\ Schmiedmayer and M.\ Oberthaler. 
\end{acknowledgments}

\pagebreak

\pagebreak
\begin{figure}
\includegraphics[width=11.2cm, angle=-90]{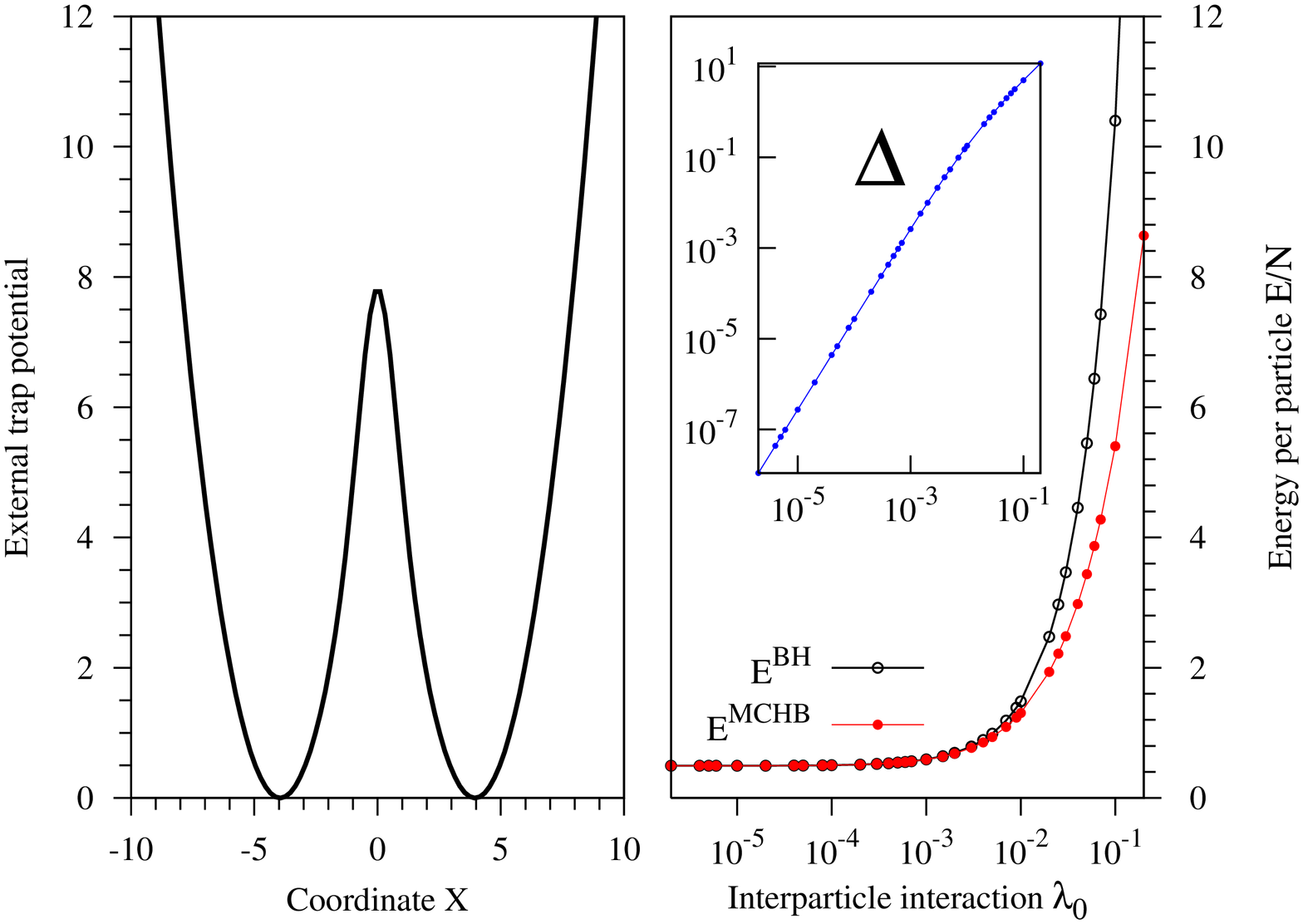}
\caption{(Color online) 
Right panel: The many-body energies per particle of the system of N=1000 trapped bosons
as a function of inter-particle interaction strength $\lambda_0$
obtained within many-body basis sets constructed on
fixed (black) and self-consistent (red) orbitals.
As the fixed orbitals we use the eigenfunctions 
of the bare Hamiltonian (BH) with potential plotted on left panel. 
The self-consistent orbitals have been obtained within the framework of the MCHB(2) method.
To demonstrate better the impact of self-consistency on the ground state energy we 
plot as $\Delta$ the difference between both energy curves in the inset. 
Left panel: 
Geometry of the symmetric one-dimensional trap used, see Eq.(\ref{Straps1D}).
It can be viewed as a combination of two harmonic traps separated by a barrier, see text for details.
}
\label{fig1}
\end{figure}

\begin{figure}
\includegraphics[width=8.2cm, angle=-90]{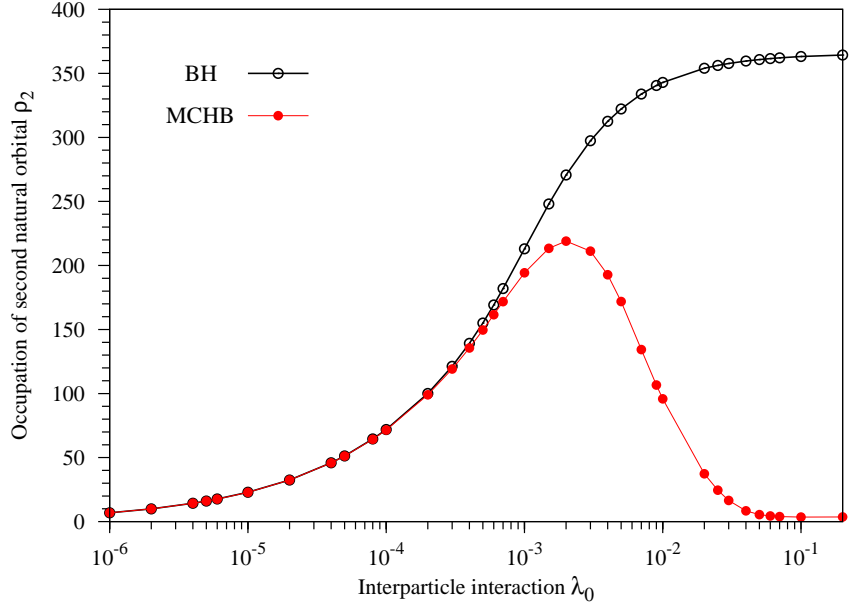}
\caption{(Color online) 
Demonstration that a lack of self-consistency can 
have a drastic impact on predicted many-body properties
of the ground state. Fragmentation is plotted as a function of the inter-particle interaction strength.
Shown is the fragmentation of $N=1000$ bosons in the symmetric double-well potential of Fig.\ref{fig1}.
The occupation number $\rho_2=N-\rho_1$ of the second natural orbital 
of the reduced one-particle density is plotted as a function of $\lambda_0$.
The solid (red) line with filled circles and the solid (black) line with open circles
mark the self-consistent (MCHB) and fixed-orbital (BH) results, respectively.
}
\label{fig2}
\end{figure}

\begin{figure}
\includegraphics[width=11.2cm, angle=-90]{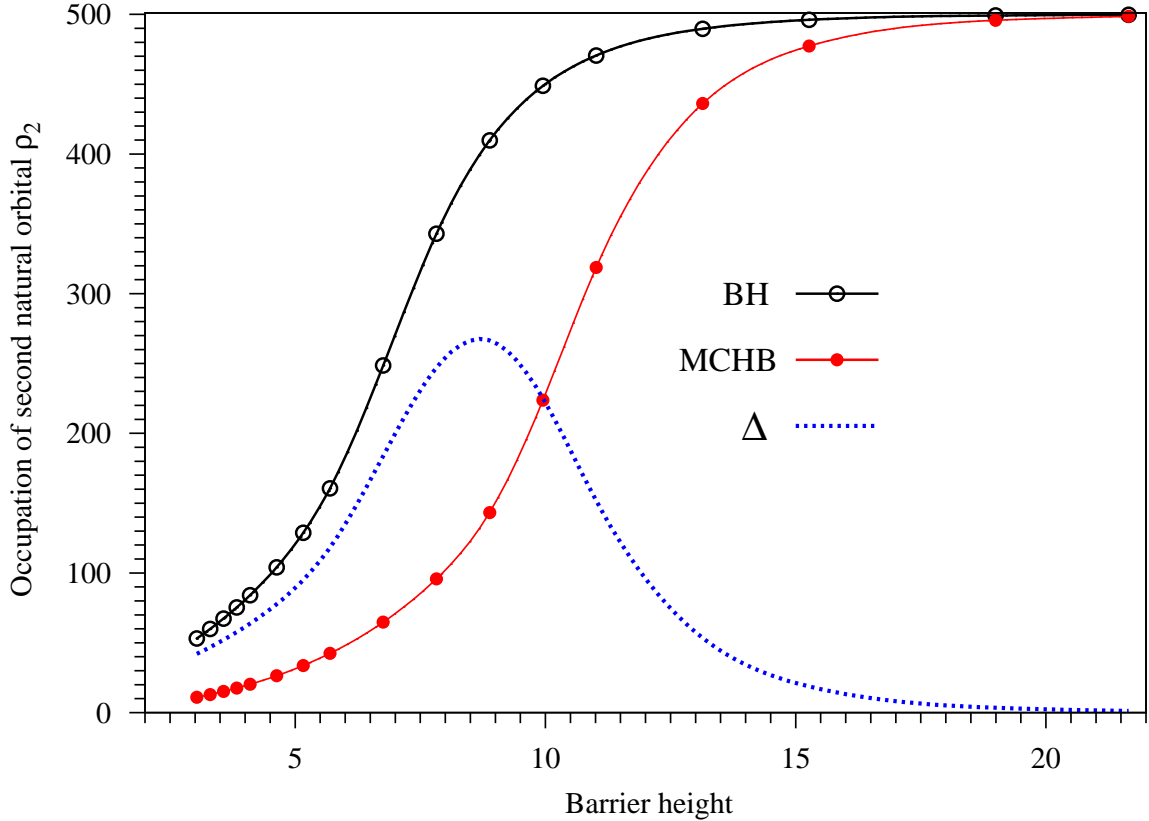}
\caption{(Color online)
Fragmentation as a function of barrier height.
The occupation number $\rho_2=N-\rho_1$ of the second natural orbital 
of the reduced one-particle density is plotted as a function of the barrier height
for N=1000 bosons trapped in a symmetric double-well potential (see text).
The solid (red) line with filled circles and the solid (black) line with open circles
represent the self-consistent (MCHB) and fixed-orbital (BH) many-body results, respectively.
To emphasize the "non-trivial" difference between self-consistent and
fixed-orbital results, the difference between both curves is plotted as a dashed (blue) line. 
The level of $\sim 25 \%$ of fragmentation is achieved at 
barrier height of $V_{symm}(0)\approx 6.5$ units with fixed-orbital many-body method
while self-consistency shifts this point to a barrier height $V_{symm}(0)\approx 10.5$ units,
see text for discussion.   
}
\label{fig3}
\end{figure}

\begin{figure}
\includegraphics[width=11.2cm, angle=-90]{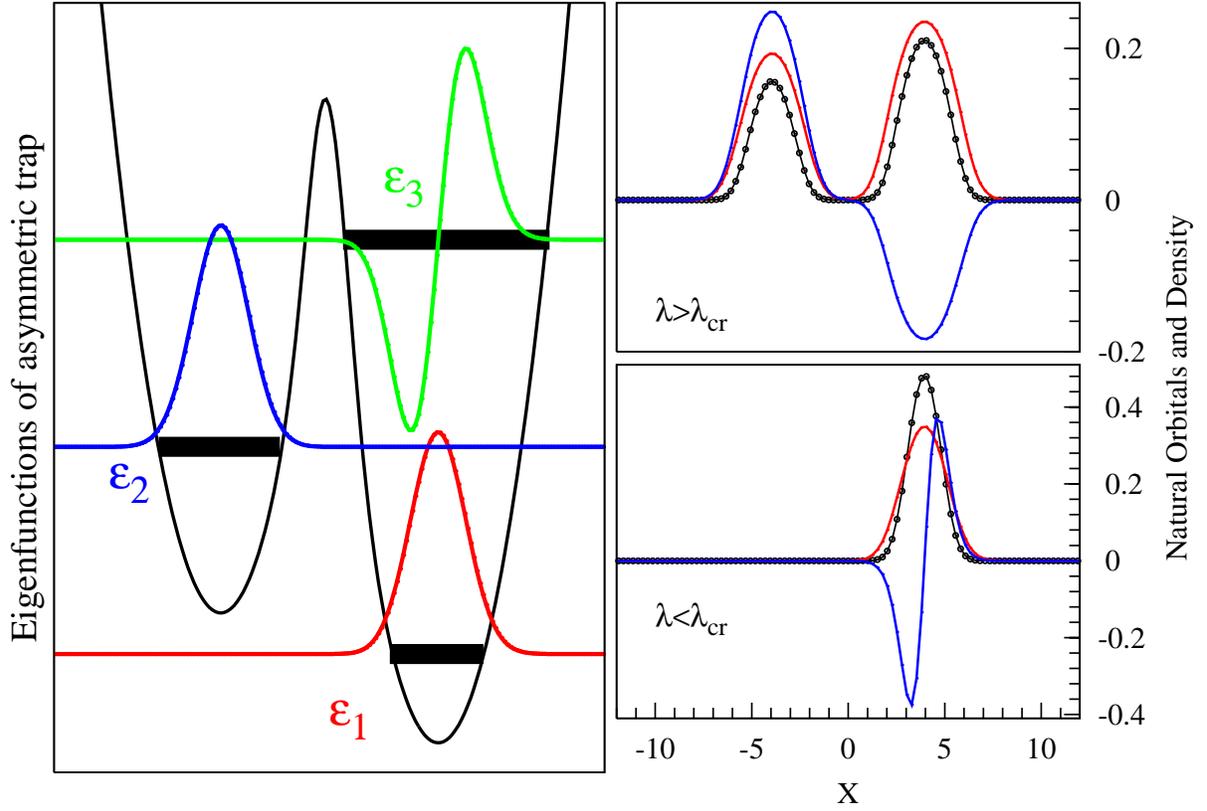}
\caption{(Color online)
Study of N=1000 bosons in an asymmetric double-well using the MCHB(2) method.
The asymmetric double-well potential and three eigenfunctions
of the respective bare Hamiltonians lowest in energy are schematically shown in the left panel.
Depending on the inter-particle interaction strength $\lambda_0$ 
the ground state of the system can enter two different regimes. 
For $\lambda<\lambda^{cr}_0$ the self-consistent MCHB(2) orbitals (solid lines) 
and respective normalized density (solid line with filled circles)
are localized in the deeper well as depicted in the right lower panel. 
In the right upper panel it is shown that for $\lambda_0>\lambda^{cr}_0$ 
the MCHB(2) natural orbitals are distributed over both wells.  
}
\label{fig5}
\end{figure}

\begin{figure}
\includegraphics[width=7.6cm, angle=-0]{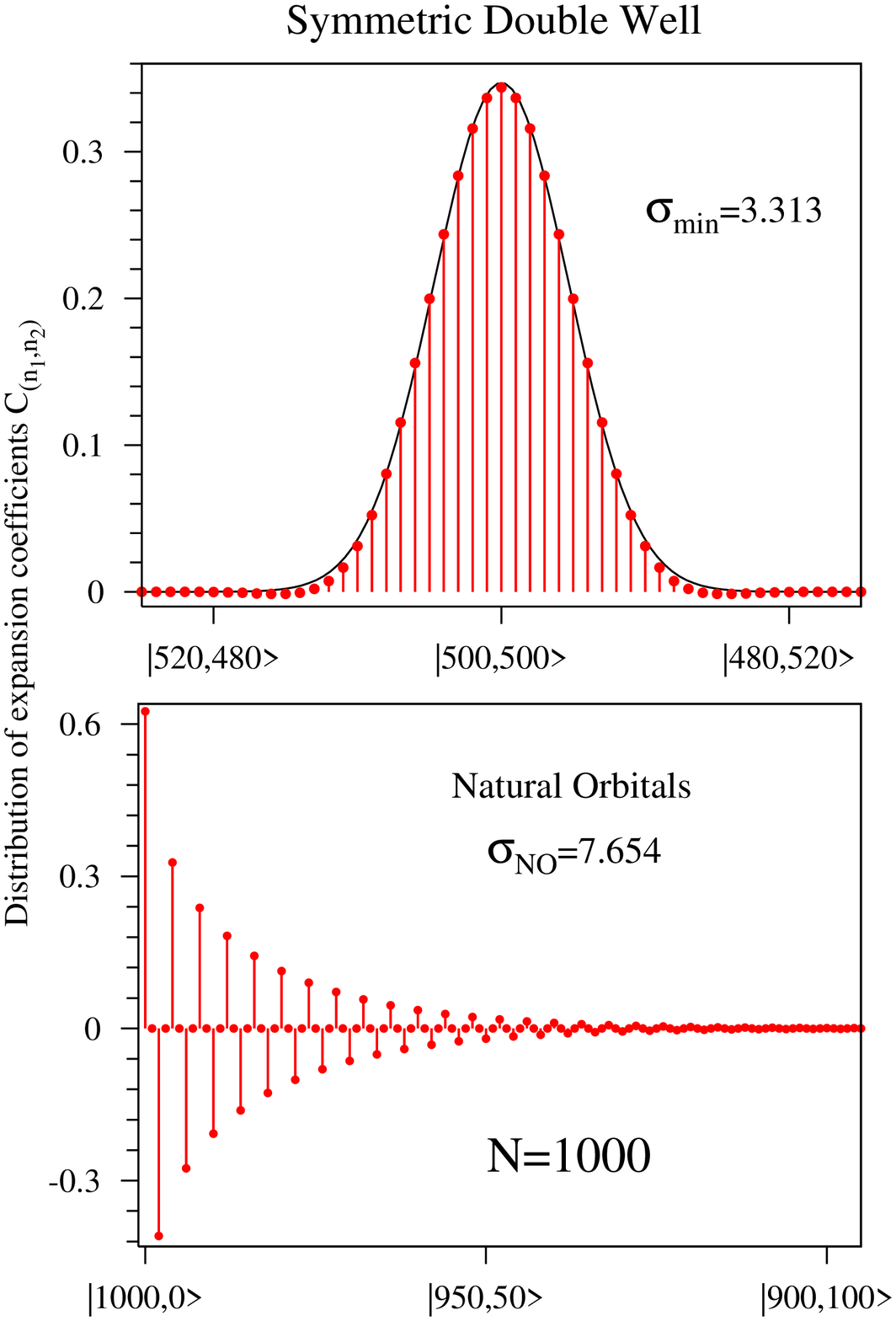}
\includegraphics[width=7.6cm, angle=-0]{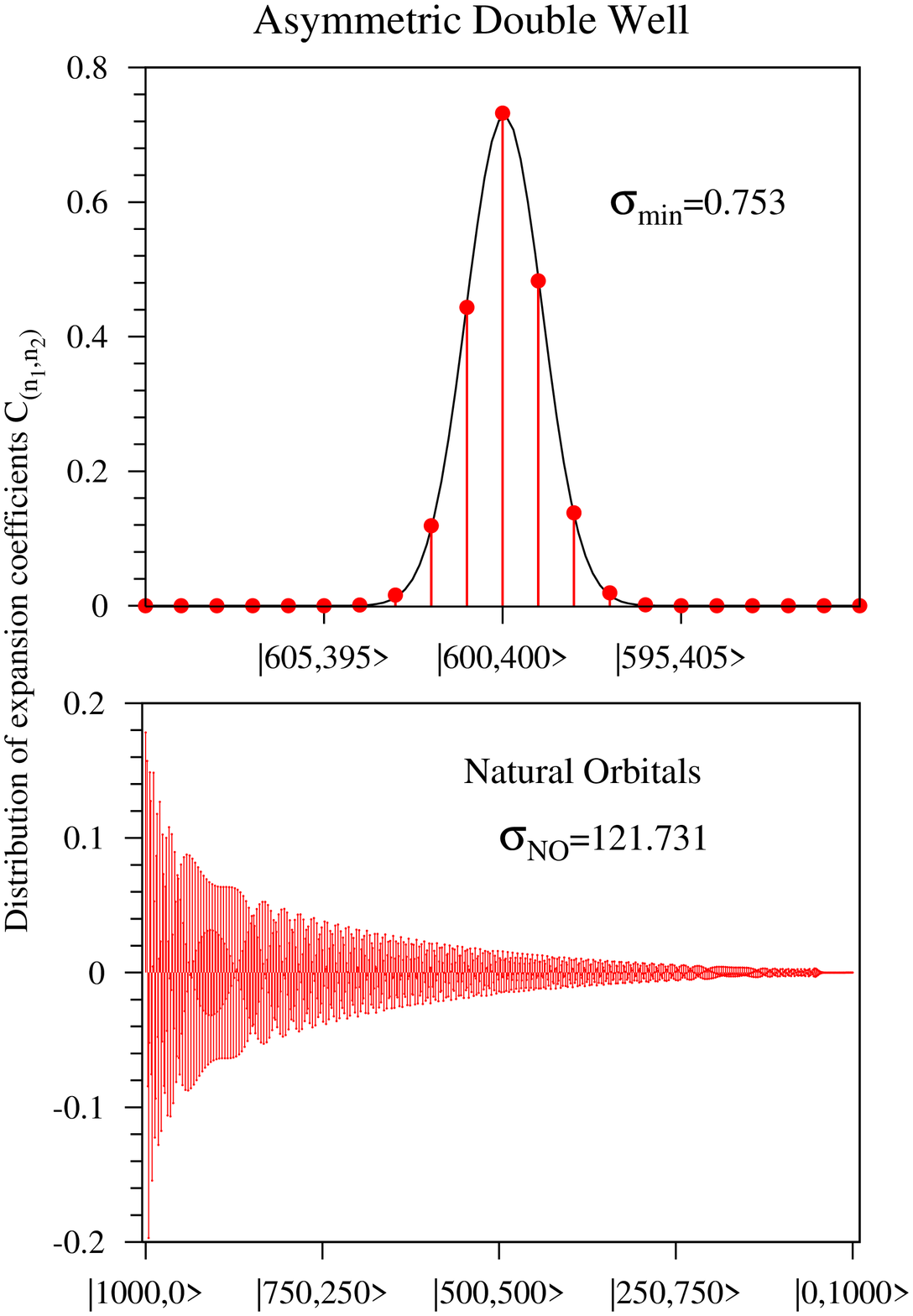}
\caption{(Color online)
The distribution of the ground state expansion coefficients $C_{(n_1,n_2)}$
obtained in MCHB(2) depends on the particular choice of the one-particle basis functions.
The left panels show the results for N=1000 bosons trapped in a symmetric double-well ($\lambda_0=0.01$).
The right panels show the analogous results for the bosons in an asymmetric double-well ($\lambda_0=0.01$).
Lower panels refer to the case where natural orbitals are used 
to construct the many-body basis functions $|n_1,n_2>$.
The width of a distribution is characterized in terms of its variance
$\sigma^2_{n_1}=\langle \hat n_1^2\rangle-\langle \hat n_1\rangle^2$.
By applying unitary transformations (rotations) on the natural orbitals, the 
width of the distributions of the expansion coefficients can be minimized.
Upper panels show the obtained distributions of the 
expansion coefficients with the minimal widths. 
The minimal distributions in these systems are well approximated by continuous Gaussian functions 
$[\frac{1}{\sigma_{min}\sqrt{2 \pi}}\exp{\frac{-(\xi-\langle n_1\rangle)^2}{2\sigma_{min}^2}}]^{1/2}$
plotted by black solid lines, see text for details.
}
\label{fig6}
\end{figure}

\begin{figure}
\includegraphics[width=11.2cm, angle=-90]{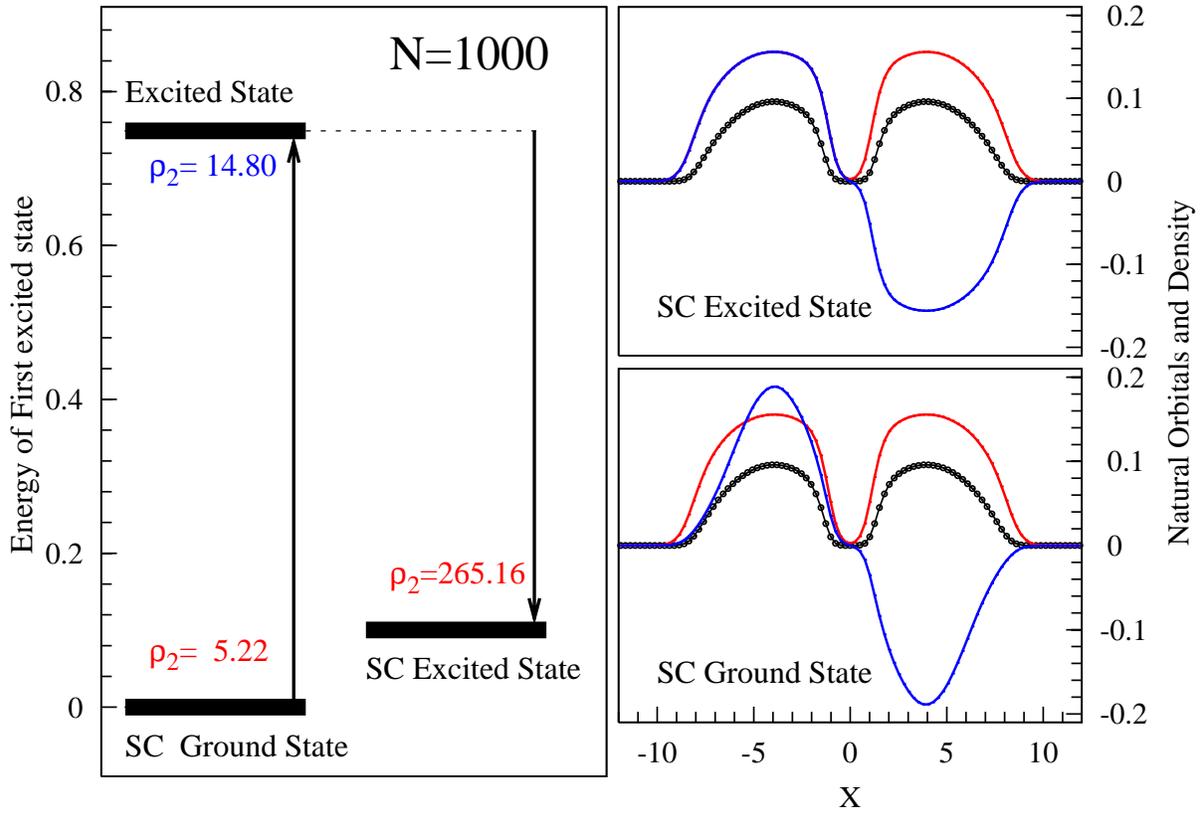}
\caption{(Color online)
Demonstration that self-consistency can have a great impact on predicted many-body properties
of excited states. Shown are results for   
N=1000 bosons trapped in a symmetric double well potential (see text for details).
Left panel: Energy levels of the ground and first excited states
obtained self-consistently are labelled as "SC Ground State" and "SC Excited State", respectively.
For comparison we present the energy level of the non-self-consistent first excited state
labelled as "Excited state", obtained by using the optimized ground state orbitals.
The occupation of the second natural orbital $\rho_2$ ($\rho_1=N-\rho_2$) is indicated for 
each state. 
Right panel: the MCHB solutions (natural orbitals) 
of the self-consistent ground and excited states (solid lines). 
The respective normalized densities are plotted by solid lines with filled circles.
}
\label{fig4}
\end{figure}

\begin{figure}
\includegraphics[width=11.2cm, angle=0]{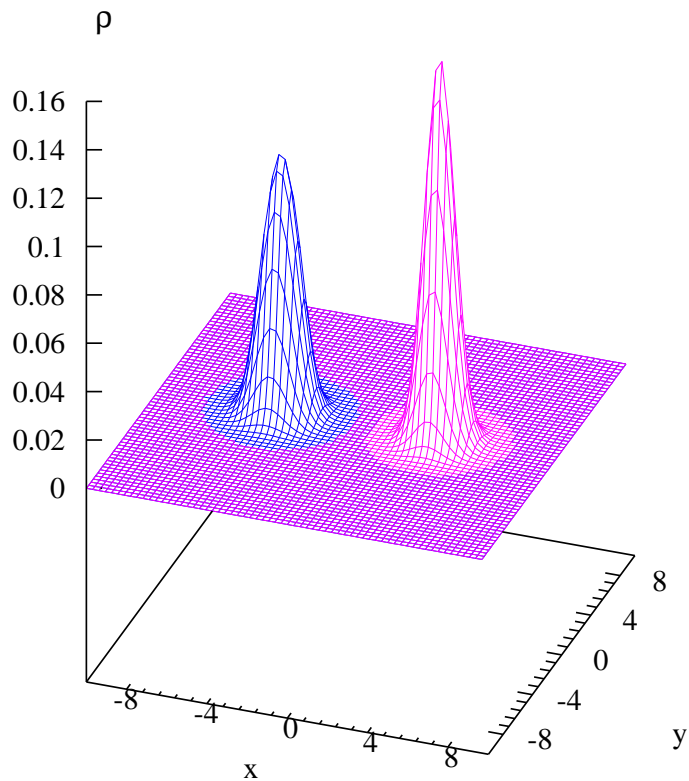}
\caption{(Color online)
Illustration of the relevance of self-consistency for
many-body treatments of bosonic systems in two-dimensions.
Ground state density of N=1000 bosons at $\lambda_0=0.01$
in the symmetric two-dimensional double-well trap of Eq.(\ref{Straps2D}).
Due to the perfect symmetry of the trap potential,
the total density consists of two equivalent parts each localized in one well.
To highlight the difference,  
the part of the self-consistent MCHB(2) ground state density localized in the left well 
is plotted together with the part of the total density obtained with the fixed-orbital many-body method
localized in the right well.
}
\label{fig2D}
\end{figure}

\begin{figure}
\includegraphics[width=8.6cm, angle=-0]{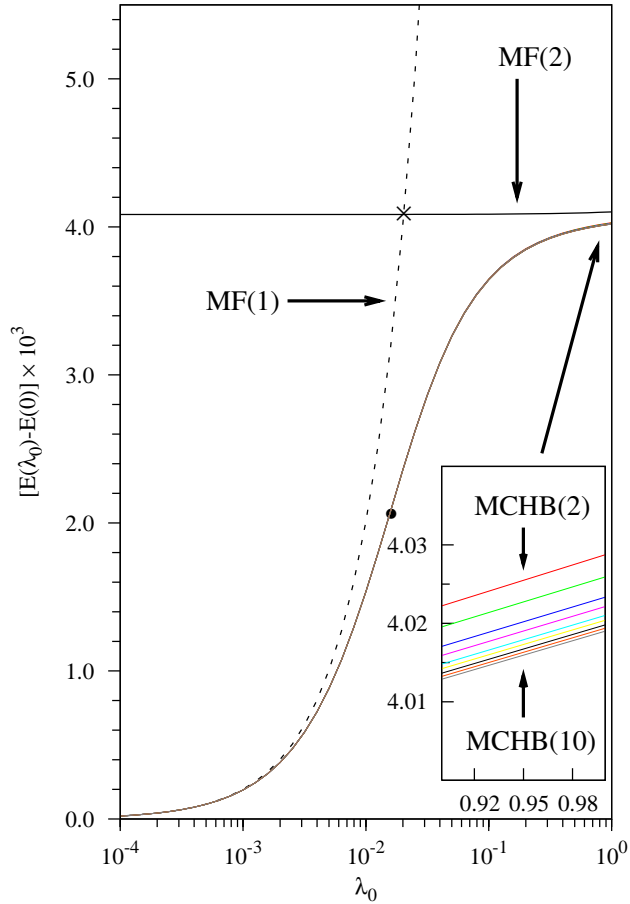}
\caption{(Color online)
The total ground state energy of two bosons trapped in a symmetric double-well potential
as a function of the inter-particle interaction strength $\lambda_0$.
All energies are plotted with respect to the ground state energy $E(0)$ of the non-interacting system.
The energy curves obtained within the framework of the two-orbital MCHB(2) and 
multi-orbital MCHB(M) M=3,$\cdots$,10 are very close to each other.
To distinguish between these curves we enlarge the scale 100 times in the inset.
To emphasize the role of many-body effects we also show the results 
obtained using the one-orbital MF(1) (Gross-Pitaevskii) and two-orbital MF(2) mean-fields.
The cross marks the transition point between "condensation" and "fragmentation" defined as 
the intersection of the MF(1) and MF(2) energy curves.
At the MCHB(M) level we observe a smooth development from "condensation" to "fragmentation"
instead of the sharp transition; the filled circle marks the respective transition point.
}
\label{fig7}
\end{figure}

\end{document}